\documentclass[prd,amsmath,amssymb,superscriptaddress,nofootinbib]{revtex4}

\newcommand{\D}{\hat{D}}

\begin{document}

\title{Effective metrics in the non-minimal
Einstein-Yang-Mills-Higgs theory}

\author{A. B. Balakin}
\email{Alexander.Balakin@ksu.ru} \affiliation{Department of
General Relativity and Gravitation, Kazan State University,
Kremlevskaya str. 18, Kazan 420008, Russia}
\author{H. Dehnen}
\email{Heinz.Dehnen@uni-konstanz.de} \affiliation{Universit\"at
Konstanz, Fachbereich Physik, Fach M 677, D-78457, Konstanz,
Germany}
\author{A. E. Zayats}%
\email{Alexei.Zayats@ksu.ru} \affiliation{Department of General
Relativity and Gravitation, Kazan State University, Kremlevskaya
str. 18, Kazan 420008,
Russia}%


\begin{abstract}

We formulate a self-consistent non-minimal five-parameter
Einstein-Yang-Mills-Higgs (EYMH) model and analyse it in terms of
effective (associated, color and color-acoustic) metrics. We use a
formalism of constitutive tensors in order to reformulate master
equations for the gauge, scalar and gravitational fields and
reconstruct in the algebraic manner the so-called associated
metrics for the Yang-Mills field. Using WKB-approximation we find
color metrics for the Yang-Mills field and color-acoustic metric
for the Higgs field in the framework of five-parameter EYMH model.
Based on explicit representation of these effective metrics for
the EYMH system with uniaxial symmetry, we consider cosmological
applications for Bianchi-I, FLRW and de Sitter models. We focus on
the analysis of the obtained expressions for velocities of
propagation of longitudinal and transversal color and
color-acoustic waves in a (quasi)vacuum interacting with
curvature; we show that curvature coupling results in time
variations of these velocities. We show, that the effective
metrics can be regular or can possess singularities depending on
the choice of the parameters of non-minimal coupling in the
cosmological models under discussion. We consider a physical
interpretation of such singularities in terms of phase velocities
of color and color-acoustic waves, using the terms ``wave
stopping'' and ``trapped surface''.

\end{abstract}

\maketitle

\section{Introduction}\label{Intro}

The coupled system of Einstein-Yang-Mills-Higgs (EYMH) equations form a
mathematical basis for the well-known self-consistent model of interaction
of gravitational, gauge and scalar fields (see, e.g., \cite{VG} for review
and basic references). The Einstein-Yang-Mills-Higgs theory unifies two
important trends in modern theory of gravity. The first one is
represented by the so-called Einstein-Yang-Mills model, which can be considered
as a non-Abelian generalization of the Einstein-Maxwell model. The second trend
is connected with the investigations of interaction of gravitational and scalar
fields. $SU(n)$ symmetric EYMH theory synthesizes the ideas and methods elaborated in
both models, inherits well-known results and presents some qualitatively new
ones. One of the most interesting results obtained in the
framework of the Einstein-Yang-Mills-Higgs model is connected with the search
for exact solutions describing non-Abelian field configurations of the monopole type
(see, e.g., \cite{2}-\cite{09}). Other important applications of the EYMH theory are
the modeling of cosmological evolution (e.g., \cite{9,10}) and modeling of particle
dynamics in the field of non-Abelian monopoles (e.g., \cite{11}). Why is
the EYMH model interesting for cosmological applications? As it was
stressed in \cite{9}, numerous models of inflation in the Early Universe involve
into consideration the multiplet of Higgs scalar fields ${\bf\Phi}$. Higgs scalar fields
are coupled with the gauge Yang-Mills field, characterized by potential
four-vectors ${\bf A}_k$, the so-called minimal coupling being realized
by means of the gauge-covariant derivative
$\hat{D}_m {\bf\Phi} = \partial_m {\bf\Phi} + [{\bf A}_m, {\bf\Phi}]$. Since contributions
of the Higgs fields into the total stress-energy tensor of the
system are considered in many models as the dominating ones at the inflation stage, the gauge
counterpart of the Higgs field, the Yang-Mills field, should also be
included into the master equations. As for the present stage of the Universe
evolution, the models with Higgs fields appear, first, in the theory of dark
matter \cite{dm,lH}, second, in the theories of dark energy \cite{DE1,DE2}.
Just the discovery of accelerated expansion of the present Universe revived an interest in
{\it non-minimal} Einstein-Yang-Mills-Higgs models in the context of search for
an explanation of the dark energy phenomenon. One of the most advanced
trends of such type is represented by the so-called $f(R)$ theories (e.g., \cite{fR}), which
are based on the appropriate (nonlinear) modifications of the
Einstein-Hilbert part of the total Lagrangian ($R/8 \pi G$) linear in the Ricci scalar.
Another version of non-minimal extension of the EYMH theory is characterized by the non-minimal
modification of a scalar field contribution to the total Lagrangian (e.g.,
\cite{Bij}). Our approach to the non-minimal modification of the EYMH theory is
presented below.

The discussion concerning the non-minimal coupling of gravity with
fields and media started at the end of the 60th. The historical
details, review and references related to the non-minimal
interaction of gravity with scalar and electromagnetic fields can
be found, e.g., in \cite{FaraR,Hehl3,BDZ}.
As for the generalization of the concept of curvature coupling for
the case of gauge field, there are two different ways to establish a
non-minimal Einstein-Yang-Mills theory. The first way is a direct
non-minimal generalization of the Einstein-Yang-Mills (EYM) theory
containing derivatives of the second order at most \cite{Horn}. In the
framework of this approach M\"uller-Hoissen obtained the non-minimal
EYM model from a dimensional reduction of the Gauss-Bonnet action
\cite{MH}. We follow the alternative way, which is connected with a
non-Abelian generalization of the non-minimal Einstein-Maxwell
theory  along the lines proposed by Drummond and Hathrell for the
linear electrodynamics \cite{Drum}. Based on the results of the
paper \cite{BL05}, we considered in \cite{1BZ06,2BZ06,BSZ07} a
three-parameter gauge-invariant non-minimal EYM model linear in
curvature.
Next natural step was a formulation of a non-minimal
Einstein-Yang-Mills-Higgs (EYMH) theory, and this process, of
course, also admits different approaches. Taking into account the
Higgs scalar multiplet  we follow, first, the ideas, proposed in
\cite{HDehnen1,HDehnen2,HDehnen3,HDehnen4}, second the concept of
derivative coupling, developed in \cite{Amen3,Capo1,Capo2}.

In this paper we establish a five-parameter non-minimal
Einstein-Yang-Mills-Higgs model. The first three coupling
parameters, $q_1$, $q_2$, $q_3$, describe a non-minimal
interaction of Yang-Mills field and gravitational field. The
fourth and fifth parameters, $q_4$, $q_5$, describe the so-called
gauge-invariant non-minimal ``derivative coupling'' of the Higgs
field with gravity. Since the gauge-invariant derivative, $\D_m
{\bf \Phi}$, contains the potential of the Yang-Mills field, the
corresponding non-minimal term is associated with ``triple''
interaction, namely, gravitational and scalar fields, gauge and
scalar fields, and gauge and gravitational fields.

It is clear that curvature induced effects predicted by the non-minimal
EYMH theory should be examined by the analysis of particle dynamics and wave
propagation.
Wave propagation in media is known to be described in the framework
of effective metric formalism (see, e.g.,
\cite{Visser1,Novello,VolovikBook} for the history details and
references). The first metric from the class of effective ones,
namely, the optical metric, was introduced by Gordon \cite{Gordon}
for isotropic media. In the effective spacetime with optical metric
light propagates as in vacuum, light rays follow geodesics
\cite{PMQ} and the wave vector is a null vector of this metric.
Written in terms of the optical metric, the constitutive equations,
linking the excitation tensor and the Maxwell tensor in
electrodynamics of isotropic media, have the same formal structure as
for vacuum electrodynamics \cite{MauginJMP,HehlObukhov}. The
second representative of the class of effective metrics is the acoustic
metric \cite{Visser1,Novello,VolovikBook}. Using this metric, one can
regard sound waves as quasi-particles moving in the effective
spacetime (e.g., \cite{Volovik}).

The covariant Fresnel equation is known to be a straightforward
way to obtain effective metrics in electrodynamics (see
\cite{HehlObukhov} and references therein). It appears as a
product of the geometrical optics approach and can be reduced to
one eikonal equation in the isotropic case. Another way is
connected with algebraic analysis of the constitutive equations.
In electrodynamics this approach is developed and successfully
used by Hehl and Obukhov \cite{HehlObukhov}. In particular, these
authors obtained the Minkowski metric as a kind of ``tensorial
square root'' of the fourth-rank constitutive tensor, linking the
excitation tensor and Maxwell (field strength) tensor in the
so-called premetric electrodynamics.

When the medium is anisotropic, light propagation is accompanied
by the phenomenon of birefringence \cite{LL}. When anisotropy is
of the uniaxial type the Fresnel equation introduces two effective
(optical) Lorentzian metrics \cite{Perlick}, which define two
light cones \cite{Visser1,Perlick,HehlObukhov}. In the biaxial
anisotropic media the effective (optical) metrics are in general
non-Lorentzian \cite{Perlick}, but again the birefringence revives
bi-metricity.

Another way (algebraic) yields the same result. As it was shown in
\cite{GRG05}, the fourth-rank constitutive tensor can
be generally reconstructed out of two symmetric second rank tensor
fields, thus introducing the so-called associated metrics. Besides, it was shown in
\cite{GRG05}, that the transition between different
representations of the constitutive tensor in terms of different pairs
of associated metrics is governed by invariance properties in an
associated two-dimensional internal vector space.
Despite the fact, that
interpretation of these associated metrics as Lorentzian optical
metrics is restricted to the uniaxial case, the representation of
constitutive equations in terms of associated metrics is valid in
linear electrodynamics of an arbitrary anisotropic medium
(exceptional case relates to the presence of skewons
\cite{HehlObukhov}).

Here we suggest the generalization of the formalism of associated
metrics for the case of gauge field and introduce a new type of
effective metrics, the color metrics. In order to motivate this
generalization, we consider here a specific sort of vacuum field
configuration, indicated as vacuum interacting with curvature, or
equivalently, non-minimal vacuum. The five-parameter non-minimal
EYMH model, considered below, is a
convenient model for our purpose due to two reasons. First,
the non-minimal EYMH model admits a clear Lagrangian formulation,
the master equations for gravitational, gauge and scalar fields
being obtained by a direct variational procedure. Second, the
non-minimal master equations look like the corresponding equations
for the anisotropic medium. In other words, in the presence of the
curvature coupling of the gravitational, gauge and scalar fields the
vacuum non-minimal EYMH model can be reformulated as a minimal one,
but in some effective medium (quasi-vacuum, or vacuum, interacting
with spacetime curvature). As a result, the effective metrics for the
Yang-Mills field (color metrics) and for the Higgs field
(color-acoustic metric) appear in a natural way.

The paper is organized as follows. In Section \ref{sec1} we
introduce basic formalism of the EYMH theory. In Section
\ref{NMEYMH} we obtain the non-minimal extensions of the master
equations for the gauge, scalar and gravity fields in the framework
of five-parameter model. In Section \ref{sec2} we discuss the
formalism of multi-metric representation of the constitutive
equations for the Yang-Mills and Higgs fields, and apply this
formalism to the case of uniaxial non-minimal vacuum. In Section
\ref{sec3} we study in detail the cosmological applications of the
effective metric formalism for Bianchi-I,
Friedmann-Lema\^itre-Robertson-Walker (FLRW) and de Sitter models.
Conclusions summarize the results. In Appendix \ref{appa} we
consider the symmetry properties of the non-minimal constitutive tensor
for the Yang-Mills and Higgs fields. Appendix \ref{appb} contains
the WKB-analysis of the problem, which explains, why the associated
metrics for the Yang-Mills field can be treated as color metrics. In
Appendix \ref{appc} we present the tidal force, which gives an
alternative description of the particle motion in the presence of
non-minimal (curvature induced) interactions.

\section{Preamble}\label{sec1}

Our aim is to consider the EYMH model with
the action functional of the following type
\begin{eqnarray}\label{0act}
S_{({\rm EYMH})} = \int d^4 x \sqrt{-g}\ \left\{ \frac{R + 2
\Lambda}{\kappa}+\frac{1}{2} C^{ikmn}_{(a)(b)}F^{(a)}_{ik}
F_{mn}^{(b)} -{\cal C}^{ik}_{(a)(b)}\hat{D}_i\Phi^{(a)}\hat{D}_k
\Phi^{(b)}+ V( \Phi^2) \right\}\,.
\end{eqnarray}
Here $g = {\rm det}(g_{ik})$ is the determinant of a metric tensor
$g_{ik}$, $R$ is the Ricci scalar, $\Lambda$ is the cosmological
constant, the multiplet of real tensor fields $F^{(a)}_{ik}$
describes the strength of gauge field, the symbol $\Phi^{(a)}$
denotes the multiplet of the Higgs scalar real fields, $\hat{D}_k$
denotes the gauge-invariant derivative, $V( \Phi^2)$ is a
potential of the Higgs field, and
$\Phi^2\equiv(\Phi^{(a)}\Phi_{(a)})$. Latin indices without
parentheses run from 0 to 3, $(a)$ and $(b)$ are the group
indices, the summation over repeating indices is assumed.
The quantities $C^{ikmn}_{(a)(b)}$ and ${\cal C}^{ik}_{(a)(b)}$ denote
the so-called constitutive tensors for the gauge and
scalar fields, respectively. They contain neither Yang-Mills
strength tensor $F^{(a)}_{ik}$, nor gauge-covariant derivative
of the Higgs field ${\D}_k \Phi^{(b)}$, i.e., the Lagrangian in
(\ref{0act}) is quadratic in these quantities.
Since the tensors
$C^{ikmn}_{(a)(b)}$ and ${\cal C}^{ik}_{(a)(b)}$ contain group
indices $(a)$ and $(b)$, the metric in the group space $G_{(a)(b)}$,
the scalar fields $\Phi^{(a)}$ and some additional directors in the
group space, $q^{(a)}$, can be used to reconstruct these tensors. The
symmetry properties of the tensors $C^{ikmn}_{(a)(b)}$ and ${\cal
C}^{ik}_{(a)(b)}$ and their general decompositions are considered in
detail in Appendix \ref{appa}. For the $U(1)$ symmetry  such theory
is discussed in \cite{BL05,AB06}, here we consider $SU(n)$ ($n>1$)
models. Let us mention that in this paper we restrict ourselves by
the condition that only the metric tensor in the group space,
$G_{(a)(b)}$, is used to construct the tensors $C^{ikmn}_{(a)(b)}$
and ${\cal C}^{ik}_{(a)(b)}$. Other constructions, for instance,
$\Phi_{(a)}\Phi_{(b)}$ tensor, etc., are considered in \cite{BDZ2}.

In addition to the real fields $F^{(a)}_{ik}$, $\Phi^{(a)}$ and
$A^{(a)}_i$ (the Yang-Mills field potential) we use below the
symbols ${\bf F}_{ik}$, ${\bf \Phi}$ and ${\bf A}_i$. In literature
there are few alternative
definitions of these quantities, despite all versions give exactly
the same result in terms of multiplets of real fields. In order to
avoid the ambiguity, we stress, that in this paper we follow the
definitions of the book \cite{Rubakov} (see Section 4.3.) Thus, we
consider the Yang-Mills field ${\bf F}_{mn}$ and the Higgs field
${\bf \Phi}$ taking values in the Lie algebra of the gauge group
$SU(n)$ (adjoint representation):
\begin{equation}
{\bf F}_{mn} = - i {\cal G} {\bf t}_{(a)} F^{(a)}_{mn} \,, \quad
{\bf A}_m = - i {\cal G} {\bf t}_{(a)} A^{(a)}_m \,, \quad {\bf
\Phi} = {\bf t}_{(a)} \Phi^{(a)} \,. \label{represent}
\end{equation}
Here ${\bf t}_{(a)}$ are the Hermitian traceless generators of
$SU(n)$ group, thus, ${\bf \Phi}$ is considered to be Hermitian,
but ${\bf F}_{mn}$ and ${\bf A}_i$ are anti-Hermitian. The group
index $(a)$ runs from $1$ to $n^2-1$. The scalar products of the
Yang-Mills and Higgs fields (indicated by bold letters)
are defined in terms of the
traces of the corresponding matrices (see, \cite{Rubakov}), the
scalar product of the generators ${\bf t}_{(a)}$ and ${\bf
t}_{(b)}$ is chosen to be equal to:
\begin{equation}
\left( {\bf t}_{(a)} , {\bf t}_{(b)} \right) \equiv 2 {\rm Tr} \
{\bf t}_{(a)} {\bf t}_{(b)}  \equiv G_{(a)(b)}  \,.
\label{scalarproduct}
\end{equation}
The symmetric tensor $G_{(a)(b)}$ plays a role of a metric in the
group space and the generators can be chosen so that the metric is
equal to the Kronecker delta. The representation
(\ref{represent}), (\ref{scalarproduct}) allows to consider the
multiplets of the real fields $\{F^{(a)}_{mn}\}$, $ \{A^{(a)}_i\}$
and $\{\Phi^{(a)} \}$ as  components of the corresponding vectors
in the $n^2-1$ dimensional group space. The operations with the
group indices $(a)$ are assumed to be the following: the repeating
indices denote the convolution, and the rule $\Phi_{(a)} =
G_{(a)(b)} \Phi^{(b)}$ for the indices lowering takes place. In
such  terms the gauge invariants in the action functional
(\ref{1act}) is reduced to
$$
{\Phi}^2\equiv \left({\bf\Phi},{\bf\Phi}\right) \Rightarrow
\Phi_{(a)} \Phi^{(a)} \,, \quad \left({\bf F}_{mn},\,{\bf
F}^{mn}\right) \Rightarrow - {\cal G}^2 F^{(a)}_{mn}F_{(a)}^{mn}
\,,
$$
\begin{equation}
\left(\D_m {\bf \Phi},\,\D^m {\bf \Phi}\right) \Rightarrow \D_m
\Phi^{(a)}\D^m \Phi_{(a)}\,. \label{reduction}
\end{equation}
The Yang-Mills fields $F^{(a)}_{mn}$ are connected with the
potentials of the gauge field $A^{(a)}_i$ by the well-known
formulas (see, e.g., \cite{Rubakov,Mosel,Akhiezer,Odin})
\begin{equation}
{\bf F}_{mn} = \nabla_m {\bf A}_n - \nabla_n {\bf A}_m + \left[ {\bf
A}_m , {\bf A}_n \right] \Rightarrow F^{(a)}_{mn} = \nabla_m
A^{(a)}_n - \nabla_n A^{(a)}_m + {\cal G} f^{(a)}_{\ (b)(c)}
A^{(b)}_m A^{(c)}_n \,. \label{Fmn}
\end{equation}
Here $\nabla _m$ is a  covariant spacetime derivative, the symbols
$f^{(a)}_{\ (b)(c)}$ denote the real structure constants of
the gauge group $SU(n)$. The gauge covariant derivative $\D_m {\bf
\Phi} \equiv {\bf t}_{(a)}\D_m \Phi^{(a)}$ is defined according to
the formulas (\cite{Rubakov}, Eqs.(4.46, 4.47))
\begin{equation}
\D_m {\bf \Phi} \equiv \nabla_m {\bf \Phi} + \left[ {\bf A}_m ,
{\bf \Phi} \right] \Rightarrow \D_m \Phi^{(a)} \equiv \nabla_m
\Phi^{(a)} + {\cal G} f^{(a)}_{\ (b)(c)} A^{(b)}_m \Phi^{(c)}
\,. \label{DPhi}
\end{equation}
For the derivative of arbitrary tensor defined in the group space
we use the following rule \cite{Akhiezer}:
\begin{eqnarray}
\D_m Q^{(a) \cdot \cdot \cdot}_{\cdot \cdot \cdot (d)} \equiv
\nabla_m Q^{(a) \cdot \cdot \cdot}_{\cdot \cdot \cdot (d)} + {\cal
G} f^{(a)}_{\cdot (b)(c)} A^{(b)}_m Q^{(c) \cdot \cdot
\cdot}_{\cdot \cdot \cdot (d)} - {\cal G} f^{(c)}_{\cdot (b)(d)}
A^{(b)}_m Q^{(a) \cdot \cdot \cdot}_{\cdot \cdot \cdot (c)} +...
\,. \label{DQ2}
\end{eqnarray}
The definition of the commutator in (\ref{Fmn}) and (\ref{DPhi})
is based on the relation
\begin{equation}
\left[ {\bf t}_{(a)} , {\bf t}_{(b)} \right] = i  f^{(c)}_{\
(a)(b)} {\bf t}_{(c)} \,, \label{fabc}
\end{equation}
providing the formula
\begin{equation}
f_{(c)(a)(b)} \equiv G_{(c)(d)} f^{(d)}_{\ (a)(b)} = - 2 i \
{\rm Tr} \ \left[ {\bf t}_{(a)} , {\bf t}_{(b)} \right] {\bf
t}_{(c)}  \,. \label{fabc1}
\end{equation}
The structure constants $f_{(a)(b)(c)}$ are supposed to be
antisymmetric under exchange of any two indices
\cite{Rubakov,Mosel,Akhiezer}. Metric $G_{(a)(b)}$ and the
structure constants $f^{(d)}_{\ (a)(c)}$ are supposed to
be constant tensors in the standard and covariant manner
\cite{Akhiezer}. This means that
\begin{equation}
\partial_m G_{(a)(b)} = 0 \,, \quad \D_m G_{(a)(b)} = 0 \,, \qquad
\partial_m  f^{(a)}_{\ (b)(c)} = 0 \,, \quad \D_m
f^{(a)}_{\ (b)(c)} = 0 \,. \label{DfG}
\end{equation}
Furthermore, when the basis ${\bf t}_{(a)}$ is chosen to provide the relation
$G_{(a)(b)} = \delta_{(a)(b)}$, it holds :
\begin{equation}
\frac{1}{n} f^{(d)}_{\ (a)(c)} f^{(c)}_{\ (d)(b)} =
\delta_{(a)(b)} = G_{(a)(b)}    \label{ff}
\end{equation}
and
\begin{equation}
\left\{ {\bf t}_{(a)} , {\bf t}_{(b)} \right\} \equiv {\bf
t}_{(a)}  {\bf t}_{(b)} + {\bf t}_{(b)} {\bf t}_{(a)} =
\frac{1}{n} \delta_{(a)(b)} {\bf I} + d^{(c)}_{\ (a)(b)} {\bf
t}_{(c)} \label{dabc}
\end{equation}
\cite{Akhiezer} with the completely symmetric coefficients
$d_{(c)(a)(b)}$ (${\bf I}$ is the matrix-unity).
The tensor $F^{ik}_{(a)}$ satisfies the relation
\begin{equation}
\hat{D}_k {}^*\! F^{ik}_{(a)} = 0 \,, \label{Aeq2}
\end{equation}
the asterisk introduces the dual tensor
\begin{equation}
{}^*\! F^{ik}_{(a)} = \frac{1}{2}\epsilon^{ikls} F_{ls (a)} \,,
\label{dual}
\end{equation}
where $\epsilon^{ikls} = \frac{1}{\sqrt{-g}} E^{ikls}$ is the
Levi-Civita tensor, $E^{ikls}$ is the completely antisymmetric
symbol with $E^{0123} = - E_{0123} = 1$.

As a first step we consider here the so-called non-minimal EYMH
model, for which the  constitutive tensors are
constructed using the spacetime and group metrics, the Riemann and
Ricci tensors, and Ricci scalar. Other possibilities will be
studied in future papers.

\section{Five-parameter non-minimal EYMH model }\label{NMEYMH}

\subsection{Non-minimal action functional}

In the context of this paper we assume, that the constitutive tensors $ C^{ikmn}_{(a)(b)}$ and
${\cal C}^{ik}_{(a)(b)}$, included into the action functional (\ref{0act}),
contain neither contributions from the gauge field ($A^{(a)}_i$, $F^{(a)}_{mn}$) nor
contributions from the Higgs field ($\Phi^{(a)}$, $\nabla_k \Phi^{(a)}$, etc.).
The  tensors $ C^{ikmn}_{(a)(b)}$ and ${\cal C}^{ik}_{(a)(b)}$
can be generally reconstructed using
tensor quantities of several types. The quantities of the first type do not contain
derivatives of the metric; they include the metric $g_{ik}$ itself,
timelike four-vector $U^k$ of the macroscopic velocity of the Yang-Mills-Higgs
system as a whole, spacelike four-vectors ${\cal D}^k_{(\alpha)}$ directed along anisotropy axes,
tensors in the group space, such as $G_{(a)(b)}$, $f^{(a)}_{\ (b)(c)}$, ...
etc. Tensor quantities of the second type contain the first derivative of the metric only,
they can be rewritten in terms of covariant derivatives $\nabla_m U^{k}$ and $\nabla_m {\cal
D}^k_{(\alpha)}$. The tensor quantities of the third type consist of second
derivatives of the metric and can be represented in terms of the Riemann tensor
$R^i_{\ kmn}$ and its convolutions: the Ricci tensor $R_{kn} \equiv R^m_{\ kmn} $ and the Ricci
scalar $R \equiv R^k_k$. The quantities of the next type may include the covariant
derivative of the Riemann tensor $\nabla_j R_{ikmn}$, the tensor $R^{ikmn} \nabla_m
U_n$, etc. When the terms containing the velocity four-vector $U^i$ and its
derivatives are taken into account, one deals with the so-called {\it
dynamo}-phenomena (see, e.g., \cite{LL} for dynamo-optical effects). When
the four-vectors ${\cal D}^k_{(\alpha)}$ and its derivatives are inserted into the
Lagrangian, one assumes that the system contains a spatially anisotropic material
subsystem. One refers to the effects induced by curvature, when the Lagrangian
contains the Riemann tensor and its convolutions, and there are no
contributions from  $U^k$ and ${\cal D}^k_{(\alpha)}$.

In this paper we focus on effects induced by curvature and consider the
Lagrangian of the EYMH theory {\it linear} in the Riemann tensor. The ansatz of linearity
with respect to Riemann tensor leads us to the five-parameter model with the action functional
written in the following (generic) form
\begin{eqnarray}\label{1act}
S_{({\rm NMEYMH})} = \int d^4 x \sqrt{-g}\ \left\{ \frac{R + 2
\Lambda}{\kappa}+\frac{1}{2}F^{(a)}_{ik} F^{ik}_{(a)}
-{\D}_m\Phi^{(a)}{\D}^m\Phi_{(a)}+ m^2 \Phi^2 \right.
{}\nonumber\\
\left. {}+\frac{1}{2} {\cal R}^{ikmn}F^{(a)}_{ik} F_{mn(a)} -
{\Re}^{\,mn}{\D}_m\Phi^{(a)}{\D}_n\Phi_{(a)} \right\}\,,
\end{eqnarray}
where the so-called susceptibility tensors ${\cal R}^{ikmn}$ and
$\Re^{\,mn}$ are defined as follows:
\begin{equation}
{\cal R}^{ikmn} \equiv \frac{q_1}{2}R\,(g^{im}g^{kn}-g^{in}g^{km})
+ \frac{q_2}{2}(R^{im}g^{kn} - R^{in}g^{km} + R^{kn}g^{im}
-R^{km}g^{in}) + q_3 R^{ikmn}\,, \label{sus}
\end{equation}
\begin{equation}\label{Reqq}
\Re^{\,mn}\equiv {q_4}Rg^{mn}+q_5 R^{mn}\,.
\end{equation}
This action describes the five-parameter non-minimal EYMH model, and
$q_1$, $q_2$, \dots\ $q_5$ are the constants of non-minimal
coupling. The action (\ref{1act}) is the particular case of
(\ref{0act}) with
\begin{equation}
C^{ikmn}_{(a)(b)} = \left[ \frac{1}{2}( g^{im} g^{kn}
- g^{in} g^{km}) + {\cal R}^{ikmn} \right] G_{(a)(b)} \,,
\label{Hik11}
\end{equation}
and
\begin{equation}\label{Heq13}
{\cal C}^{mn}_{(a)(b)} = G_{(a)(b)} \left( g^{mn} + \Re^{\,mn}
\right) \,.
\end{equation}
Below we indicate the tensor
\begin{equation}
\tilde{g}^{ik}= g^{ik} + \Re^{ik}
\label{cam}
\end{equation}
as a color-acoustic metric.

\vspace{3mm}

\noindent
{\it REMARK 1}

\noindent
The non-minimal susceptibility tensors ${\cal R}^{ikmn}$ (\ref{sus}) and $\Re^{\,mn}$ (\ref{Reqq}) can be
rewritten in terms of irreducible parts of the curvature tensor
\begin{equation}\label{irredu}
{\cal R}^{ikmn} {=} \lambda_1{\cal G}^{ikmn} {+} \lambda_2 {\cal E}^{ikmn}
{+}  \lambda_3 {\cal C}^{ikmn}  \,, \quad \Re^{\,mn} \equiv  \lambda_4 Rg^{mn} {+} \lambda_5 {\cal S}^{mn}
\end{equation}
where ${\cal C}^{ikmn}$ is the traceless Weyl tensor
\begin{equation}\label{irredu3}
{\cal C}^{ikmn} \equiv  R^{ikmn} {-} {\cal E}^{ikmn}
{-} {\cal G}^{ikmn}  \,, \quad {\cal C}^{m}_{\ \ nmk} =0 \,,
\end{equation}
and
$$
{\cal E}^{ikmn} \equiv \frac{1}{2} \left({\cal S}^{im}g^{kn} {-} {\cal S}^{in}g^{km} {+} {\cal S}^{kn}g^{im}
{-} {\cal S}^{km}g^{in}) \right) \,, \quad {\cal S}^{mn} \equiv R^{mn}
{-} \frac{1}{4} R g^{mn} \,,
$$
\begin{equation}\label{irredu2}
{\cal G}^{ikmn} \equiv \frac{1}{12} R \ (g^{im}g^{kn} {-} g^{in}g^{km}) \,,
\end{equation}
(we use the notations from the book \cite{ExactSol}). The phenomenological
parameters $\lambda_1$, $\lambda_2$, $\lambda_3$, $\lambda_4$ and
$\lambda_5$ are connected  with $q_1$, $q_2$, $q_3$, $q_4$ and $q_5$
by the linear relations
\begin{equation}\label{irredu7}
\lambda_1 = 6q_1 {+} 3q_2 {+} q_3 \,, \quad \lambda_2 = q_2 {+} q_3 \,, \quad \lambda_3 = q_3\,,
\quad  \lambda_4  = q_4 {+} \frac{1}{4} q_5 \,,  \quad \lambda_5 = q_5 \,.
\end{equation}
The representations (\ref{sus}) and (\ref{Reqq})  are equivalent
to (\ref{irredu}) from a computational point of view,
nevertheless, we use below the decomposition (\ref{sus}) taking
into account some historical motivation (see, e.g., reviews
\cite{FaraR,Hehl3}). In particular, the phenomenological model
proposed by Prasanna in \cite{Prasanna1} contains the Riemann
tensor only ($q_1=q_2=q_4=q_5=0$) and gives a first example of the
violation of the strong equivalence principle in the framework of
non-minimal electrodynamics. The first non-phenomenological model
elaborated by Drummond and Hathrell in \cite{Drum} is based on
one-loop corrections to quantum electrodynamics in curved
spacetime; the coupling parameters are calculated directly in
\cite{Drum} yielding $q_1 =-5q$, $q_2=13q$, $q_3=-2q$ (or
equivalently $\lambda_1= 7q$, $\lambda_2 = 11q$, $\lambda_3=-2q$).
The positive parameter $q$, $q \equiv \frac{\alpha\lambda^2_{\rm
e}}{180\pi}$ is constructed by using the fine structure constant
$\alpha$, and the Compton wavelength of the electron $\lambda_{\rm
e}$. The models investigated by Horndeski \cite{Horn,Horn1}, by
M\"uller-Hoissen \cite{MH}, by M\"uller-Hoissen and Sippel
\cite{MHS}, contain master equations of the second order only;
they are characterized by the following values of the coupling
parameters: $q_1 = - q$, $q_2= 2q$, $q_3=-q$ (or equivalently
$\lambda_1= -q$, $\lambda_2 = q$, $\lambda_3=-q$) (here parameter
$q$ is arbitrary). These relations can be recovered by the ansatz
that the non-minimal susceptibility tensor ${\cal R}_{ikmn}$ is
proportional to the double dual Riemann tensor $^{*}R^{*}_{ikmn}$,
i.e., ${\cal R}_{ikmn} = \gamma\; ^{*}R^{*}_{ikmn}$, for some
$\gamma$ \cite{BL05}. Analogously, one can use the Weyl tensor
${\cal C}_{ikmn}$ in the relation ${\cal R}_{ikmn} = \omega\,{\cal
C}_{ikmn}$, for some $\omega$ \cite{BL05}. Thus, there are
different physical and mathematical motives to specify coupling
constants, but here we prefer to consider generic $q_1$, $q_2$,
... $q_5$, introduced phenomenologically.

\vspace{3mm}

\noindent
{\it REMARK 2}

\noindent
The Lagrangian of the non-minimal model linear in curvature can contain also the term
$\xi R \Phi^{(a)} \Phi_{(a)}$ without gradients of the Higgs fields, the corresponding model
with $\xi \neq 0$ is the six-parameter one (see, e.g., \cite{BDZ}).
Here we focus on five-parameter EYMH model only.

\vspace{3mm}

\noindent
{\it REMARK 3}

\noindent
Let us summarize the assumptions, which we made in the formulation of the
five-parameter EYMH model.

\noindent
{\it (i)}
The Lagrangian of the EYMH model is quadratic in $F^{(a)}_{mn}$ and $\hat{D}_m
\Phi^{(a)}$ (see (\ref{0act})).

\noindent {\it (ii)} The Lagrangian of the EYMH model is linear in
the curvature tensor (see (\ref{1act})-(\ref{Reqq})).

\noindent
{\it (iii)}
The group indices in (\ref{1act}) are coupled by the metric $G_{(a)(b)}$
only.

The latter assumption leads to an important consequence: it provides all the constitutive tensors to be
multiplicative, i.e., they are products of the metric in the group space and the corresponding tensors
defined in the spacetime. In its turn, this consequence guarantees, that for the five-parameter EYMH model the
constitutive tensors are symmetric with respect to transposition of the group indices.

\subsection{Non-minimal extension of the Yang-Mills
equations}

The variation of the action $S_{({\rm NMEYMH})}$ with respect to
the Yang-Mills potential $A^{(a)}_i$ yields
\begin{equation}
\D_k {\cal H}^{ik}_{(a)}  =  - {\cal G} (\D_k
\Phi^{(b)})f_{(a)(b)(c)} \Phi^{(c)} \left( g^{ik} + \Re^{ik}
\right) \,, \label{Heqs}
\end{equation}
where the tensor ${\cal H}^{ik}_{(a)}$  is defined as
\begin{equation}
{\cal H}^{ik}_{(a)} = \left[ \frac{1}{2}( g^{im} g^{kn} - g^{in}
g^{km}) + {\cal R}^{ikmn} \right] G_{(a)(b)} F^{(b)}_{mn} \,.
\label{HikR}
\end{equation}
Equivalently, one can write
\begin{equation}
{\D}_k {\bf H}^{ik} \equiv \nabla_k {\bf H}^{ik}+\left[{\bf
A}_k,{\bf H}^{ik}\right] = {\cal
G}^2\left[\left(g^{ik}+\Re^{\,ik}\right)\D_k{\bf
\Phi},{\bf\Phi}\right] \,, \label{YMeq}
\end{equation}
where
\begin{equation}
{\bf H}^{ik} = {\bf F}^{ik} + {\cal R}^{ikmn} {\bf F}_{mn} \,.
\label{33YMeq}
\end{equation}
By analogy with electrodynamics of continuous media the tensor
${\bf H}^{ik}$ in the non-minimal Yang-Mills equations can be
called excitation tensor. Below we use the term
``color excitation''. Equation (\ref{HikR}) is in fact a linear
constitutive law, connecting the color excitation tensor $
H^{ik}_{(a)}$ and the field strength tensor $F^{(b)}_{mn}$ (see,
e.g., \cite{Mauginbook,LL,HehlObukhov} for the $U(1)$ symmetry).
It can be clearly rewritten as
\begin{equation}
{\cal H}^{ik}_{(a)} = C^{ikmn}_{(a)(b)} \ F^{(b)}_{mn} \,,
\label{Hik}
\end{equation}
demonstrating that the quantity ${\cal R}^{ikmn}G_{(a)(b)}$ plays the role of
color susceptibility tensor. The term ${\cal G}^2
\left(g^{ik}+\Re^{\,ik}\right) \left[\D_k{\bf
\Phi},{\bf\Phi}\right]$ plays, respectively, the role of color
current induced by the Higgs fields.

\subsection{Non-minimal extension of the Higgs field
equations}

The variation of the action $S_{({\rm NMEYMH})}$ with respect to
the Higgs scalar field $\Phi^{(a)}$ yields
\begin{equation}\label{Heq}
{\D}_m\left[ \left( g^{mn} + \Re^{\,mn} \right)\D_n{\bf
\Phi}\right] = - m^2 {\bf\Phi} \,.
\end{equation}
By analogy with thermodynamics the four-vector $\D_n{\bf \Phi}$
can be indicated as color four-gradient of the Higgs field (the
analog of the temperature three-gradient), and the four-vector
${\bf \Psi}^m$ can be called the color flux four-vector (the
analog of the heat flux). Thus, the relation
\begin{equation}\label{Heq11}
{\bf \Psi}^m =  \left( g^{mn} + \Re^{\,mn} \right)\D_n{\bf \Phi}
\end{equation}
is in fact a non-minimal constitutive law, generalizing the
Fourier law in thermodynamics, its general form being
\begin{equation}\label{Heq12}
\Psi^m_{(a)} = {\cal C}^{mn}_{(a)(b)}  (\D_n{\bf \Phi})^{(b)} \,.
\end{equation}

\subsection{Master equations for the gravitational
field}

In the non-minimal theory linear in curvature, the equations for
the gravity field related to the action functional $S_{({\rm
NMEYMH})}$ take the form
\begin{equation}
\left(R_{ik}-\frac{1}{2}Rg_{ik}\right)= \Lambda g_{ik} +
{\kappa}T^{(NMYMH)}_{ik} \,. \label{Eeq}
\end{equation}
The principal novelty of these equations in comparison with the
well-known equations for non-minimal scalar field is associated
with the third, fourth, etc., terms in the decomposition
\begin{equation}
T^{(NMYMH)}_{ik} =  T^{(YM)}_{ik} + T^{(H)}_{ik} + q_1
T^{(I)}_{ik} + q_2 T^{(II)}_{ik} + q_3 T^{(III)}_{ik} + q_4
T^{(IV)}_{ik} + q_5 T^{(V)}_{ik} \,. \label{Tdecomp}
\end{equation}
The first term $T^{(YM)}_{ik}$:
\begin{equation}
T^{(YM)}_{ik} \equiv \frac{1}{4} g_{ik} F^{(a)}_{mn}F^{mn}_{(a)} -
F^{(a)}_{in}F_{k\,(a)}^{\ n} \,, \label{TYM}
\end{equation}
is a stress-energy tensor of pure Yang-Mills field. The second
one, $T^{(H)}_{ik}$,
\begin{equation}
T^{(H)}_{ik}=\D_i\Phi^{(a)}\D_k\Phi_{(a)}-\frac{1}{2}g_{ik}\D_m\Phi^{(a)}\D^m\Phi_{(a)}+\frac{1}{2}m^2
\Phi^2 \,g_{ik}
\end{equation}
is a stress-energy tensor of the Higgs field. The definitions of
other five tensors are related to the corresponding coupling
constants $q_1$, $q_2,\,\dots,\,q_5$:
\begin{equation}%
T^{(I)}_{ik} = R\,T^{(YM)}_{ik} -  \frac{1}{2} R_{ik}
F^{(a)}_{mn}F^{mn}_{(a)} + \frac{1}{2} \left[ {\D}_{i} {\D}_{k} -
g_{ik} {\D}^l {\D}_l \right] \left[F^{(a)}_{mn}F^{mn}_{(a)}
\right] \,, \label{TI}
\end{equation}%
\[%
T^{(II)}_{ik} = -\frac{1}{2}g_{ik}\biggl[{\D}_{m}
{\D}_{l}\left(F^{mn(a)}F^{l}_{\ n(a)}\right)-R_{lm}F^{mn (a)}
F^{l}_{\ n(a)} \biggr] \]%
\[{}- F^{ln(a)}
\left(R_{il}F_{kn(a)} + R_{kl}F_{in(a)}\right)-R^{mn}F^{(a)}_{im}
F_{kn(a)} - \frac{1}{2} {\D}^m{\D}_m \left(F^{(a)}_{in}
F_{k\,(a)}^{ \
n}\right) \]%
\begin{equation}%
\quad{}+\frac{1}{2}{\D}_l \left[ {\D}_i \left(
F^{(a)}_{kn}F^{ln}_{(a)} \right) + {\D}_k
\left(F^{(a)}_{in}F^{ln}_{(a)} \right) \right] \,, \label{TII}
\end{equation}%
\[
T^{(III)}_{ik} = \frac{1}{4}g_{ik} R^{mnls}F^{(a)}_{mn}F_{ls(a)}-
\frac{3}{4} F^{ls(a)} \left(F_{i\,(a)}^{\ n} R_{knls} +
F_{k\,(a)}^{\ n}R_{inls}\right) \]%
\begin{equation}%
\quad {}-\frac{1}{2}{\D}_{m} {\D}_{n} \left[ F_{i}^{ \ n
(a)}F_{k\,(a)}^{ \ m} + F_{k}^{ \ n(a)} F_{i\,(a)}^{ \ m} \right]
\,, \label{TIII}
\end{equation}%

\[%
T^{(IV)}_{ik}=\left(R_{ik}-\frac{1}{2}Rg_{ik}\right)\D_m\Phi^{(a)}\D^m\Phi_{(a)}+R\,\D_i\Phi^{(a)}\D_k\Phi_{(a)}
\]%
\begin{equation}\label{TIV}
    {}+\left(g_{ik}\D_n\D^n-\D_i\D_k\right)\left[\D_m\Phi^{(a)}\D^m\Phi_{(a)}\right]\,,
\end{equation}

\[%
T^{(V)}_{ik}=\D_m\Phi^{(a)}\left[R_i^m\D_k\Phi_{(a)}+R_k^m\D_i\Phi_{(a)}\right]-
\frac{1}{2}R_{ik}\,\D_m\Phi^{(a)}\D^m\Phi_{(a)}
\]%
\[%
{}+\frac{1}{2}g_{ik}\D_m\D_n\left[\D^m\Phi^{(a)}\D^n\Phi_{(a)}\right]
\]%
\begin{equation}\label{TV}
    {}-\frac{1}{2}\,\D^m\biggl\{\D_i\left[\D_m\Phi^{(a)}\D_k\Phi_{(a)}\right]+
\D_k\left[\D_m\Phi^{(a)}\D_i\Phi_{(a)}\right]-
    \D_m\left[\D_i\Phi^{(a)}\D_k\Phi_{(a)}\right]\biggr\}\,.
\end{equation}
The Einstein tensor $G_{ik} \equiv R_{ik}-\frac{1}{2}g_{ik}R$ is
the divergence-free one, thus, the tensor $T^{(NMYMH)}_{ik}$ in
the right-hand-side of (\ref{Eeq}) has to satisfy the differential
condition
\begin{equation}
\nabla^k T^{(NMYMH)}_{ik} =0 \,. \label{Eeeq}
\end{equation}
One can prove that it is valid automatically, when $F^{(a)}_{ik}$
is a solution of the Yang-Mills equations (\ref{YMeq}), and
$\Phi^{(a)}$ satisfy the Higgs equations (\ref{Heq}). In order to
check this fact directly, one has to use the Bianchi identities
and the properties of the Riemann tensor:
\begin{equation}
\nabla_i R_{klmn} + \nabla_l R_{ikmn} + \nabla_k R_{limn} = 0 \,,
\quad R_{klmn} + R_{mkln} + R_{lmkn} = 0 \,, \label{bianchi}
\end{equation}
as well as the rules for the commutation of covariant derivatives
\begin{equation}
(\nabla_l \nabla_k - \nabla_k \nabla_l) {\cal A}^i = {\cal A}^m
R^i_{\cdot mlk} \,, \label{nana}
\end{equation}
(this rule is written here for vectors only). The procedure of
checking is analogous to the one, described in \cite{Acci3} and we
omit it here.

\subsection{Interpretation of the gravity field equations in terms of
gravitational excitation and strength}

Since the Riemann tensor $R^{i}_{\ klm}$ can be considered as a
gravitational field strength, an analog of $F_{ik}$ in electrodynamics
(see, e.g., \cite{MTW}), one can emphasize an interesting analogy
with electrodynamics. Let us mention that one of the ingredients of
electrodynamics of media is a linear constitutive law
\begin{equation}
H^{im} = {\cal H}^{im} + C^{imls} F_{ls} \,, \label{Z2}
\end{equation}
where $H^{im}$ is an excitation tensor, $C^{imls}$ is a
constitutive tensor and ${\cal H}^{im}$ is a spontaneous
polarization-magnetization tensor, which does not depend on the Maxwell
tensor \cite{Mauginbook}. The Maxwell tensor and the excitation
tensor satisfy equations
\begin{equation} \nabla_{i} F_{kl} +
\nabla_{l} F_{ik} + \nabla_{k} F_{li} =0\,, \quad  \nabla_k H^{ik}
= - \frac{4\pi}{c} I^{i} \,,
\label{Z44}
\end{equation}
where $I^{i}$ is a four-vector of the current of unbounded
charges. In analogy with (\ref{Z2}) we can introduce a linear
constitutive equation for the gravity field
\begin{equation}
Z_{imkn} = {\cal H}_{imkn} + {\cal C}_{imkn}^{\ \ \ \ \ \ pqls}
R_{pqls} \,, \label{Z1}
\end{equation}
where $Z_{imkn}$ is a tensor of gravitational excitation, ${\cal
C}_{imkn}^{\ \ \ \ \ pqls}$ is a  constitutive tensor,
and ${\cal H}_{imkl}$ is a tensor of gravitational polarization.
In analogy with (\ref{Z44}) the field strength $R_{pqls}$
satisfies the Bianchi identity (\ref{bianchi}). The Einstein equations
(\ref{Eeq}) with (\ref{Tdecomp}) look like the convolution of the
constitutive equations (\ref{Z1}) with $g^{mn}$, if we put
$$
{\cal C}_{imkn}^{\ \ \ \ \ \ pqls} = \frac{1}{2}\left [{\cal
L}_{ik}^{ \ \ ([pq][ls])} g_{mn} - {\cal L}_{in}^{\ \ ([pq][ls])}
g_{mk} + {\cal L}_{mn}^{\ \ ([pq][ls])} g_{ik} - {\cal L}_{mk}^{\
\ ([pq][ls])} g_{in} \right]
$$
$$
{}- \frac{1}{6} \left( g_{ik}g_{mn} - g_{in}g_{mk} \right)
g^{rt}{\cal L}_{rt}^{ \ \ ([pq][ls])} \,,
$$
$$
{\cal L}_{ik}^{\ \ pqls} \equiv g^{qs} \left[\delta^p_i \delta^l_k
- \frac{1}{2}g^{pl} g_{ik} \right] + \frac{3}{4} \kappa q_3\left[
F^{ls}_{(a)} \left(\delta^{p}_{i} F_{k}^{{(a)}
q}+\delta^{p}_{k}F_{i}^{{(a)} q} \right) - \frac{1}{3}g_{ik}
F^{pq}_{(a)}F^{ls {(a)}} \right]
$$
$$
{}+ \kappa g^{qs} \left\{ \frac{1}{2}\delta^p_i \delta^l_k \left[
q_1 F^{(a)}_{jh}F^{jh}_{(a)} + \left( q_5 - 2
q_4\right)\D_j\Phi^{(a)}\D^j\Phi_{(a)} \right] \right.
$$
$$
\left. {}- g^{pl} \left[ q_1 T^{(YM)}_{ik} + q_4 \left(
T^{(H)}_{ik} - \frac{1}{2} m^2 \Phi^2 g_{ik}\right) \right]- q_5
\D^{l}\Phi^{(a)} \left[\delta^{p}_{i}\D_{k}\Phi_{(a)} +
\delta^{p}_{k}\D_{i}\Phi_{(a)} \right] \right.
$$
\begin{equation}
\left. + q_2 \left[- \frac{1}{2} g_{ik} F^{p {(a)}}_{\ \ j}
F^{lj}_{(a)} + F^{lj}_{(a)} \left(\delta^p_i F^{(a)}_{kj} +
\delta^p_k F^{(a)}_{ij} \right) + F_{i{(a)}}^{\ p} F_k^{{(a)} l}
\right] \right\} \label{Z3}
\end{equation}
for the constitutive tensor, and
\begin{equation}
{\cal H}_{imkn}=\kappa\left({\cal H}^{(G)}_{i[n}g_{k]m}-{\cal
H}^{(G)}_{m[n}g_{k]i}- \frac{1}{3} {\cal H}^{(G)}_{ls}g^{ls}
g_{i[n}g_{k]m}\right)\,,
\label{Z4}
\end{equation}
$$
{\cal H}^{(G)}_{ik} = \Lambda g_{ik} + T^{(YM)}_{ik} +
T^{(H)}_{ik} + \frac{1}{2}q_1 \left[ {\D}_{i} {\D}_{k} - g_{ik}
{\D}^l {\D}_l \right] \left[F^{(a)}_{mn}F^{mn}_{(a)} \right]
$$
$$
- \frac{1}{2}q_2 \left\{ g_{ik}\left[{\D}_{m}
{\D}_{l}\left(F^{mn(a)}F^{l}_{\ n(a)}\right) \right] +{\D}^m{\D}_m
\left(F^{(a)}_{in} F_{k\,(a)}^{ \ n}\right) \right.
$$
$$
\left. - {\D}_l \left[ {\D}_i \left( F^{(a)}_{kn}F^{ln}_{(a)}
\right) + {\D}_k \left(F^{(a)}_{in}F^{ln}_{(a)} \right) \right]
\right\}
$$
$$
-\frac{1}{2}q_3{\D}_{m} {\D}_{n} \left[ F_{i}^{ \ n
(a)}F_{k\,(a)}^{ \ m} + F_{k}^{ \ n(a)} F_{i\,(a)}^{ \ m} \right]
+ q_4
\left(g_{ik}\D_n\D^n-\D_i\D_k\right)\left[\D_m\Phi^{(a)}\D^m\Phi_{(a)}\right]
$$
$$
- \frac{1}{2} q_5 \left\{ \,\D^m
\left\{\D_i\left[\D_m\Phi^{(a)}\D_k\Phi_{(a)}\right]+\D_k\left[\D_m\Phi^{(a)}\D_i\Phi_{(a)}\right]
    - \D_m\left[\D_i\Phi^{(a)}\D_k\Phi_{(a)}\right] \right\} \right.
$$
\begin{equation}
\left. - g_{ik}\D_m\D_n\left[\D^m\Phi^{(a)}\D^n\Phi_{(a)}\right]
\right\}
 \label{Z5}
\end{equation}
for the gravitational excitation. Here we use the following standard definitions for symmetrization
and antisymmetrization of indices, respectively:
\begin{equation}
{\cal K}_{(mn)} \equiv \frac{1}{2} \left({\cal K}_{mn} {+} {\cal K}_{nm}  \right) \,, \quad
{\cal K}_{[mn]} \equiv \frac{1}{2} \left({\cal K}_{mn} {-} {\cal K}_{nm}
\right)\,.
 \label{symantisym}
\end{equation}
Finally, the identity
\begin{equation}
\nabla^k \left\{ R_{ik} -\frac{1}{2}g_{ik} g^{mn}R_{mn} - \kappa
T^{(NMYMH)}_{ik} - \Lambda g_{ik}\right\} =0 \,,
 \label{Z7}
\end{equation}
written as
\begin{equation}
\nabla^k \left(Z_{imkn} g^{mn} \right) =0 \,,
 \label{Z77}
\end{equation}
is an analog of the second Maxwell equation (\ref{Z44}) with
vanishing current. Clearly, the constitutive tensor (\ref{Z3})
entering the constitutive law (\ref{Z1}), is built using metric,
quadratic combinations of the tensor $F^{(a)}_{ik}$ and color
gradient of the Higgs field, and does not contain the derivatives of
the spacetime metric.

Thus, in the presented five-parameter non-minimal model three
constitutive tensors, $C^{ikmn}_{(a)(b)}$ (\ref{Hik11}), ${\cal
C}^{ik}_{(a)(b)}$ (\ref{Heq13}) and ${\cal C}_{imkn}^{\ \ \
\ \ \ \ pqls}$ (\ref{Z3}), appear in a natural way. It can be
considered as a motivation for the construction of more
sophisticated models with the action functional (\ref{0act}).

\section{Multi-metric representation of the constitutive
tensors}\label{sec2}

\subsection{General formalism}

Our final purpose is to reconstruct effective metrics (associated, color and
color-acoustic) for the five-parameter EYMH model with the action functional
(\ref{1act}) for the one-axis Bianchi-I model, FLRW and de Sitter models. In order to explain
the procedure of their finding for this five-parameter model, we consider, first,
a general formalism  (see next subsection and Apendices), then we reduce basic
formulas to the case of uniaxial symmetry, and, finally, we obtain desired effective metrics.
Let us mention, that we do not introduce new assumptions in addition to
the three ones fixed above in the Remark 3, but we made a few
simplifications, which follow from the symmetry of the corresponding cosmological model.
For instance, the spacetime symmetry of the one-axis Bianchi-I model requires the symmetry
of the constitutive tensors to be uni-axial with vanishing cross-effects.
This is a reason, why we consider the appropriate reduced formulas instead
of the general ones.

\subsubsection{Decomposition of the tensor $C^{ikmn}_{(a)(b)} $}

Let $g^{im(\alpha)}_{(a)}$ be tensor fields symmetric with respect
to spacetime indices, $i$ and $m$. Here $\alpha = 1,2$ is the
field number, and $(a)$ is the group index (color number), taking
only one value for $U(1)$ symmetry and running from $1$ to $n^2-1$
for the $SU(n)$ symmetry. This Latin index is placed in
parentheses. We restrict ourselves by the models with
$C^{ikmn}_{(a)(b)} = C^{ikmn}_{(b)(a)}$, thus, the
most general representation of this constitutive tensor in terms of
$g^{im(\alpha)}_{(a)}$  is
\begin{equation}
C^{ikmn}_{(a)(b)} =
\frac{1}{2\hat{\mu}}\sum_{(\alpha)(\beta){(c)(d)}}
 G_{(\alpha)(\beta)(a)(b)}^{(c)(d)}\left(g^{im (\alpha)}_{(c)} \ g^{kn
(\beta)}_{(d)} - g^{in (\alpha)}_{(c)} \ g^{km (\beta)}_{(d)}
\right) \,, \label{supergeneral}
\end{equation}
where the following symmetry of indices is assumed:
\begin{equation}
G_{(\alpha)(\beta)(a)(b)}^{(c)(d)} =
G_{(\alpha)(\beta)(b)(a)}^{(c)(d)} =
G_{(\beta)(\alpha)(a)(b)}^{(d)(c)}
 \,, \label{s22}
\end{equation}
the sum being over all possible combinations of field numbers and
color numbers. The factor $\hat{\mu}$ is introduced, as in
\cite{GRG05}, for convenience. The $C^{ikmn}_{(a)(b)}$ tensor has
the general form (\ref{Cdecomp}) (see Appendix \ref{appa}). It
contains $21 \frac{(n^2-1)n^2}{2}$ components. There are 10
components of $g^{im (\alpha)}_{(c)}$ for each $\alpha$ and $(a)$.
Thus, when we deal with two metrics, i.e., $\alpha=1,2$ for $n>1$
one has $20(n^2-1)$ field components $g^{im(\alpha)}_{(a)}$, as
well as $\frac{1}{2}(n^2-1)^2 n^2(2n^2-1)$ parameters (functions)
$G_{(\alpha)(\beta){(a)(b)}}^{(c)(d)}$ , as a tool for
reconstruction of $21 \frac{(n^2-1)n^2}{2}$ components of the
$C^{ikmn}_{(a)(b)}$ tensor. The reconstruction of
$C^{ikmn}_{(a)(b)}$ along the line of (\ref{supergeneral}) is always
possible since the corresponding algebraic system is
underdetermined. Let us mention that $U(1)$ case is a special one, and
(see, e.g., \cite{GRG05}) one has $20$ field components
$g^{im(\alpha)}$, as well as 3 parameters $G_{(\alpha)(\beta)}$ to
reconstruct 21 components of $C^{ikmn}$. In the case of $SU(2)$
symmetry we have, generally, $60$ field components
$g^{im(\alpha)}_{(a)}$, 126 components of
$G^{(c)(d)}_{(\alpha)(\beta)(a)(b)}$ and 126 components of
$C^{ikmn}_{(a)(b)}$. Thus, 60 parameters are arbitrary.

The quantity $G_{(\alpha)(\beta){(a)(b)}}^{(c)(d)}$ is considered
to be a twice covariant and twice contravariant tensor in the
$n^2-1$ dimensional group space, the tensors $g^{im
(\alpha)}_{(c)}$ are the vectors in that space. When $\alpha$, $i$
and $m$ are fixed, the standard unitary transformations are
assumed to conserve the scalar product $G^{(a)(b)}g^{im
(\alpha)}_{(a)}g^{kn (\beta)}_{(b)}$. As well, when the spacetime
indices and group indices are fixed, there exist a linear
transformation from one set of $g^{im (\alpha)}_{(c)}$ to another one,
and $g^{im (\alpha)}_{(c)}$ can be regarded as vectors in some
effective two-dimensional space \cite{GRG05}. When we deal with
the so-called ``parallel fields'' \cite{Yasskin}, i.e., when
\begin{equation}
g^{im (\alpha)}_{(a)} = q_{(a)} g^{im (\alpha)} \,, \quad
 G^{(a)(b)}q_{(a)}q_{(b)} = 1 \,,
\label{parall}
\end{equation}
we can put
\begin{equation}
G_{(\alpha)(\beta){(a)(b)}}^{(c)(d)} = G_{(\alpha)(\beta)}
G_{(a)(b)} G^{(c)(d)} \,, \label{multi}
\end{equation}
providing $C^{ikmn}_{(a)(b)}$ to have a multiplicative structure,
$C^{ikmn}_{(a)(b)} = C^{ikmn} G_{(a)(b)}$. Thus, the model of
parallel fields $g^{im (\alpha)}_{(a)}$ for the $SU(n)$ case
covers the $U(1)$ model considered in \cite{GRG05}. When, on the
contrary, we assume $C^{ikmn}_{(a)(b)} = C^{ikmn} G_{(a)(b)}$, the
tensor $G_{(\alpha)(\beta){(a)(b)}}^{(c)(d)}$ is not obligatory
multiplicative, nevertheless, the parallel fields  give one of the
solutions of the reconstruction problem.

As any symmetric tensor, the quantities $g^{ik(\alpha)}_{(a)}$ can
be decomposed with respect to their components parallel and
orthogonal to the four-velocity $U^{i}$,
\begin{equation}
g^{ik (\alpha)}_{(a)} = {\cal B}^{(\alpha)}_{(a)} U^i U^k + {\cal
D}^{i (\alpha)}_{(a)} U^k + {\cal D}^{k (\alpha)}_{(a)} U^i +
{\cal S}^{ik (\alpha)}_{(a)} \,, \label{gviaU}
\end{equation}
where
\begin{equation}
{\cal B}^{(\alpha)}_{(a)} = g^{ik (\alpha)}_{(a)}U_i U_k \,, \quad
{\cal D}^{p (\alpha)}_{(a)} \equiv  \Delta^p_i
g^{ik(\alpha)}_{(a)} U_k \,, \qquad {\cal S}^{pq (\alpha)}_{(a)}
\equiv \Delta^p_i g^{ik(\alpha)}_{(a)} \Delta^q_k \,.
\label{gviaUD}
\end{equation}
In the application of this formalism
to cosmology (see below), one obtains directly that
\begin{equation}
{\cal D}^{p (\alpha)}_{(a)}
= 0  \quad \Rightarrow \quad g^{ik (\alpha)}_{(a)} =  U^i U^k + {\cal S}^{ik (\alpha)}_{(a)}
\quad \Rightarrow \quad U_ig^{im (\alpha)}_{(a)} = U^{m} \,,
\label{gviaU1}
\end{equation}
which means that the velocity four-vector $U^k$ is the eigenvector
for all tensors $g^{ik (\alpha)}_{(a)}$ with the eigenvalue equal
to one.

\subsubsection{Decomposition of the tensor ${\cal
C}^{ik}_{(a)(b)}$}

We assume that ${\cal C}^{ik}_{(a)(b)}={\cal C}^{ik}_{(b)(a)} = {\cal C}^{ki}_{(b)(a)}$
and consider the following decomposition based on the effective metrics
$\tilde{g}^{ik}_{(a)}$
\begin{equation}
{\cal C}^{ik}_{(a)(b)} = \Gamma^{(c)}_{(a)(b)}
\tilde{g}^{ik}_{(c)}\,. \label{aca1}
\end{equation}
The tensor ${\cal C}^{ik}_{(a)(b)}$ is multiplicative, i.e.,
${\cal C}^{ik}_{(a)(b)}= {\cal C}^{ik} G_{(a)(b)}$, and one can choose
tensor $\tilde{g}^{ik}_{(c)}$ independent on group index and put
\begin{equation}
{\cal C}^{ik}_{(a)(b)} = G_{(a)(b)} \tilde{g}^{ik} \,.
\label{aca2}
\end{equation}
This choice can be motivated by the case of parallel fields
\begin{equation}
\tilde{g}^{ik}_{(c)} = \tilde{g}^{ik} \ q_{(c)} \,, \label{aca3}
\end{equation}
for which the following representation is possible
\begin{equation}
\Gamma^{(c)}_{(a)(b)} q_{(c)} = G_{(a)(b)} \,. \label{aca4}
\end{equation}
The most general decomposition can be much more sophisticated, and
we do not consider it here.

\subsection{Effective metrics for the five-parameter EYMH model}
\subsubsection{General case}

In the framework of the five-parameter non-minimal EYMH model the
macroscopic tensor has multiplicative structure, i.e.,
$C^{ikmn}_{(a)(b)}= C^{ikmn} G_{(a)(b)}$, the color metrics
$g^{ik(\alpha)}_{(a)}$ can be chosen independent on the group
indices. One can say that in this case there is no ``group
multi-refringence'', but the spacetime birefringence exists. Then,
following \cite{GRG05} we represent the tensor $C^{ikmn}_{(a)(b)}$
(\ref{Hik11}) by the decomposition
$$
C^{ikmn}_{(a)(b)} = G_{(a)(b)} \frac{1}{2 \hat{\mu}} \left\{
\left[g^{im (A)} g^{kn (A)} - g^{in (A)} g^{km (A)}\right] \right.
$$
\begin{equation}
\left. - \gamma \left[(g^{im (A)}-g^{im (B)})(g^{kn (A)}-g^{kn
(B)}) - (g^{in (A)}-g^{in (B)})(g^{km (A)}-g^{km (B)}) \right]
\right\} \,, \label{maindecomp}
\end{equation}
by choosing
\begin{equation}
G_{(A)(A)} + G_{(B)(A)} = 1\ ,\quad G_{(A)(B)} + G_{(B)(B)} = 0
\,, \quad \gamma \equiv G_{(A)(B)} = G_{(B)(A)}
 \,. \label{ConditionsG}
\end{equation}
Here the indices $(A)$ and $(B)$ are no longer arbitrary: they
are fixed in order to simplify the decomposition of the constitutive
tensor \cite{GRG05}. We will refer below to this choice as to
A-metric, $g^{im (A)}$, and B-metric, $g^{im (B)}$, as well as to
the corresponding A-wave and B-wave. It is reasonable since
generally the A-wave and B-wave are not obligatory ``ordinary''
and ``extraordinary'', as in classical optics.

With this decomposition the A-tensor, $g^{ik(A)}$, can be readily
written as (see \cite{GRG05} for details)
\begin{equation}
 g^{ik(A)} = U^iU^k + \hat{\mu} \varepsilon^{ik} \,. \label{gA}
\end{equation}
The B-tensor, $g^{ik(B)}$, can be represented in terms of
eigenvectors $X^i_{(1)},X^i_{(2)}$, and $X^i_{(3)}$  of the
permittivity tensor  $\varepsilon^{ik}$ \cite{Perlick}. These
eigenvectors form the triad and possess the properties
\begin{equation}
g_{ik} X^i_{(\alpha)} X^k_{(\beta)} = - \delta_{(\alpha)(\beta)} ,
\quad g_{ik} X^i_{(\alpha)} U^k = 0 \,, \label{eigen1}
\end{equation}
\begin{equation}
X^i_{(1)}X^k_{(1)} + X^i_{(2)}X^k_{(2)} + X^i_{(3)}X^k_{(3)} = -
g^{ik} + U^i U^k = - \Delta^{ik}\,, \label{eigen2}
\end{equation}
\begin{equation}
\epsilon^{i}_{ \ klm}U^m X^k_{(\alpha)}X^l_{(\beta)} =
\epsilon_{(\alpha)(\beta)(\gamma)} X^i_{(\gamma)}, \label{eigen21}
\end{equation}
where $(\alpha), (\beta).... = (1), (2), (3)$ are tetrad indices;
there is a summation over repeating indices. The
permittivity tensor can be decomposed according to
\begin{equation}
\varepsilon^{ik} = - \sum_{(\alpha)}\varepsilon_{(\alpha)}
X^i_{(\alpha)} X^k_{(\alpha)}, \quad \varepsilon^{k}_k =
\varepsilon_{(1)} + \varepsilon_{(2)} + \varepsilon_{(3)},
\label{exx}
\end{equation}
where the terms $\varepsilon_{(\alpha)}$ denote the eigenvalues,
corresponding to the eigenvector $X^i_{(\alpha)}$. Since the tensor
$\varepsilon^{ik}$ is orthogonal to the four-velocity vector $U^i$,
the corresponding eigenvalue $\varepsilon_{(0)}$ is equal to zero,
and the velocity does not appear in this decomposition.

The structure of the B-tensor, $g^{im (B)}$, is much more sophisticated
in the general case. Keeping in mind an application of this model to
cosmology, we present here only one example, when the
tetrad components of the impermeability tensor
$(\mu^{-1})_{(\alpha)(\beta)}$ are diagonal, i.e.,
\begin{equation}
(\mu^{-1})_{(\alpha)(\beta)} = \frac{1}{\mu_{(\alpha)}}
\delta_{(\alpha)(\beta)} \,, \label{mudiag}
\end{equation}
and the tensor of cross-effects vanishes.
In this case the B-tensor (see \cite{GRG05}) has the form
\begin{equation}
g^{im (B)} = g^{im (A)} \mp \hat{\mu} \sqrt{\frac{M_{(1)} M_{(2)}
M_{(3)}}{\gamma}} \sum_{(\alpha)= (1) }^{(3)}
\frac{\varepsilon_{(\alpha)}}{M_{(\alpha)}} X^i_{(\alpha)}
X^m_{(\alpha)} \,, \label{gBgeneral}
\end{equation}
where
\begin{equation}
M_{(\alpha)} \equiv 1 -
\frac{\varepsilon_{(\alpha)}}{\mu_{(\alpha)}
\hat{\mu}\varepsilon_{(1)}\varepsilon_{(2)} \varepsilon_{(3)} } .
\label{Ma}
\end{equation}

\subsubsection{Uniaxial case}

Below we consider the applications of the model to the spacetimes
with uniaxial three-dimensional subspaces and simplify
the structure of the A- and B-tensors accordingly to this case.
Let the direction of the privilege axis be $Ox^3$ and
$M_{(1)}=M_{(2)} = 0$. Then one obtains
\begin{equation}\label{Bepsilon}
\varepsilon_{\bot} \equiv \varepsilon^{1}_{1} =
\varepsilon^{2}_{2}
 = 1 + 2 {\cal R}^{1t}_{ \ \ 1t}  \,, \quad
\varepsilon^{3}_{3} \equiv \varepsilon_{||} = 1 + 2 {\cal
R}^{3t}_{ \ \ 3t} \,,
\end{equation}
\begin{equation}\label{Bmu}
\frac{1}{\mu_{\bot}} \equiv (\mu^{-1})^{1}_{1} =
(\mu^{-1})^{2}_{2}
 = 1 + 2 {\cal R}^{23}_{ \ \ 23}\,, \quad
(\mu^{-1})^{3}_{3} \equiv \frac{1}{\mu_{||}} = 1 + 2 {\cal
R}^{12}_{ \ \ 12} \,,
\end{equation}
\begin{equation}\label{Bg111}
\frac{1}{\hat{\mu}} = \varepsilon_{||} \varepsilon_{\bot}
\mu_{\bot} \,, \quad \frac{1}{\gamma} = 1 -
\frac{\varepsilon_{||}\mu_{\bot}}{\varepsilon_{\bot}\mu_{||}} \,.
\end{equation}
The A- and B-tensors take the form
\begin{equation}\label{Bg1u}
g^{ik (A)} = U^{i} U^{k} - \frac{1}{\varepsilon_{||} \mu_{\bot}}
\left( X^{i}_{(1)}X^{k}_{(1)} + X^{i}_{(2)} X^{k}_{(2)}\right) -
\frac{1}{\varepsilon_{\bot} \mu_{\bot}} X^{i}_{(3)} X^{k}_{(3)}
\,,
\end{equation}
\begin{equation}\label{Bg2u}
g^{ik (B)} = U^{i} U^{k} - \frac{1}{\varepsilon_{\bot} \mu_{||}}
\left( X^{i}_{(1)}X^{k}_{(1)} + X^{i}_{(2)} X^{k}_{(2)}\right) -
\frac{1}{\varepsilon_{\bot} \mu_{\bot}} X^{i}_{(3)} X^{k}_{(3)}
\,,
\end{equation}
providing the relation (\ref{maindecomp}). In other words, they
give the associated metrics for the non-minimal EYMH model.
In the WKB-approximation the Yang-Mills equations are
satisfied when
\begin{equation}\label{Bas}
g^{ik (A)} p_i p_k = 0 \,, \quad \hbox{or} \quad g^{ik (B)} p_i
p_k =0 \,,
\end{equation}
(see Appendix \ref{appb}). Thus, these associated metrics can be
interpreted as the color ones. This means, in particular, that the
propagation of a test Yang-Mills wave in the non-minimally active
spacetime can be described by two dispersion relations
\begin{equation}\label{om1}
\omega^2_{(A)} =  \frac{p^2_{\bot}}{\varepsilon_{||} \mu_{\bot} }
+ \frac{p^2_{||}}{\varepsilon_{\bot} \mu_{\bot}} \,, \quad
\omega^2_{(B)} =  \frac{p^2_{\bot}}{\varepsilon_{\bot} \mu_{||} }
+ \frac{p^2_{||}}{\varepsilon_{\bot} \mu_{\bot}} \,,
\end{equation}
where
\begin{equation}\label{om2}
\omega = p_l U^l  \,, \quad p^2_{\bot} = - (p_1 p^1 + p_2 p^2) \,,
\quad p^2_{||} = - p_3 p^3 \,.
\end{equation}
When the wave propagates in the longitudinal direction, i.e.,
$p_{\bot}=0$, or in the transverse one, i.e., $p_{||}=0$, the
corresponding phase velocities can be defined as
\begin{equation}\label{om300}
{\cal V}_{||} \equiv \frac{\omega}{p_{||}} \,, \quad  {\cal
V}_{\bot} \equiv \frac{\omega}{p_{\bot}} \,.
\end{equation}
For A-wave and B-wave they are, respectively,
$$
{\cal V}^{(A)}_{||} = {\cal V}^{(B)}_{||} =
\frac{1}{\sqrt{\varepsilon_{\bot}\mu_{\bot}}} =
\sqrt{\frac{1+2{\cal R}^{23}_{\ \ 23}}{1+2{\cal R}^{1t}_{\ \ 1t}}}
\,,
$$
\begin{equation}\label{om3}
{\cal V}^{(A)}_{\bot} =\frac{1}{\sqrt{\varepsilon_{||}\mu_{\bot}}} =
\sqrt{\frac{1+2{\cal R}^{23}_{\ \ 23}}{1+2{\cal R}^{3t}_{\ \ 3t}}} \,,
\quad {\cal V}^{(B)}_{\bot} =
\frac{1}{\sqrt{\varepsilon_{\bot}\mu_{||}}} = \sqrt{\frac{1+2{\cal R}^{12}_{\ \ 12}}{1+2{\cal R}^{1t}_{\ \ 1t}}} \,.
\end{equation}
These quantities depend on spacetime metric, and its derivatives
and convert into one, when the non-minimal coupling constants
vanish, i.e., $q_1=q_2=q_3=0$ (let us repeat, that in units used
here the speed of light in the standard vacuum is equal to one).
The color-acoustic metric can be represented in the form analogous to
(\ref{Bg1u}):
\begin{equation}\label{acumet11}
\tilde{g}^{ik} = {\cal A}^2 \left[ U^{i} U^{k} - \frac{1}{\sigma^2_{\bot}}
\left( X^{i}_{(1)}X^{k}_{(1)} + X^{i}_{(2)} X^{k}_{(2)}\right) -
\frac{1}{\sigma^2_{||}} X^{i}_{(3)} X^{k}_{(3)} \right]
\,,
\end{equation}
where
\begin{equation}\label{acumet111}
{\cal A}^2 = 1+{\cal R}^{t}_{t} \,, \quad
\sigma_{\bot} = \sqrt{\frac{1+{\cal R}^{t}_{t}}{1+{\cal R}^{1}_{1}}} \,,
\quad
\sigma_{||} = \sqrt{\frac{1+{\cal R}^{t}_{t}}{1+{\cal R}^{3}_{3}}}
\,,
\end{equation}
and $\sigma_{\bot}$ and $\sigma_{||}$ can be interpreted as phase velocities of a transversal and
longitudinal waves, respectively.

\section{Cosmological application of the effective metric
formalism}\label{sec3}

\subsection{One-axis Bianchi-I model}

Relativistic cosmology needs effective metric paradigm since it
uses observational data obtained by analysis of the directional
distribution and frequency properties of incoming photons.
Propagating photons are influenced by numerous regular and
stochastic phenomena, and we could try to identify this influence,
if we reconstruct the effective dielectric permittivity and magnetic
permeability of the regions, where the photons passed. Effective
metric formalism gives mathematical grounds for such a
reconstruction.

Let us use the metric
\begin{equation}\label{Bianchi}
ds^2 = dt^2 - \left[ a^2(t) \left(dx^2 + dy^2 \right) + c^2(t)
dz^2 \right] \,,
\end{equation}
which is attributed to the well-known Bianchi-I cosmological model
with two equivalent spatial directions, $Ox$ and $Oy$. In
\cite{BZ05} we presented the exact solutions of the non-minimally
extended Einstein-Maxwell model with such metric in case when
there is a magnetic field directed along $0z$. That model can be
automatically generalized to the case of parallel Yang-Mills field
$F^{(a)}_{ik}= q^{(a)} B (\delta^{1}_{i} \delta^{2}_{k} -
\delta^{2}_{i} \delta^{1}_{k})$ along the line proposed in
\cite{Yasskin}. Thus, we can refer to the exact solutions,
published in \cite{BZ05}, as to the ones with non-minimally
coupled parallel Yang-Mills field of the magnetic type. This means
that we can use both terms: optical metric and color metric
referring  to that solutions.
The tetrad vectors for this metric are
\begin{equation}\label{Btetrad}
X^k_{(0)}= \delta^k_0 \,, \quad X^k_{(1)}= \delta^k_1
\frac{1}{a(t)} \,, \quad X^k_{(2)}= \delta^k_2 \frac{1}{a(t)} \,,
\quad X^k_{(3)}= \delta^k_3 \frac{1}{c(t)}  \,.
\end{equation}
Based on the symmetry of the spacetime we easily obtain that the
non-minimal susceptibility tensors ${\cal R}^{ik}_{ \ \ mn}$ and
$\Re^m_n$ correspond to the case of uniaxial symmetry. The
non-vanishing components of these tensors are the following
\begin{equation}\label{R1}
{-} {\cal R}^{1t}_{ \ \ 1t} {=} \frac{1}{2}(4q_1 {+} 3q_2 {+}
2q_3)\frac{\ddot{a}}{a} {+} \frac{1}{2}(2q_1 {+} q_2)
\frac{\ddot{c}}{c} {+} \frac{1}{2}(2q_1 {+}
q_2)\frac{\dot{a}^2}{a^2} {+} \frac{1}{2}(4q_1 {+}
q_2)\frac{\dot{a}}{a}\frac{\dot{c}}{c} \,,
\end{equation}
\begin{equation}\label{R4}
{-} {\cal R}^{13}_{ \ \ 13} {=} \frac{1}{2}(4q_1 {+}
q_2)\frac{\ddot{a}}{a} {+} \frac{1}{2}(2q_1 {+}
q_2)\frac{\ddot{c}}{c} {+} \frac{1}{2}(2q_1 {+}
q_2)\frac{\dot{a}^2}{a^2} {+} \frac{1}{2}(4q_1 {+} 3q_2 {+}
2q_3)\frac{\dot{a}}{a}\frac{\dot{c}}{c} \,,
\end{equation}
\begin{equation}\label{R2}
{-} {\cal R}^{3t}_{ \ \ 3t} {=} (2q_1 {+} q_2)\frac{\ddot{a}}{a}
{+} (q_1 {+} q_2{+}q_3)\frac{\ddot{c}}{c} {+} q_1
\frac{\dot{a}^2}{a^2} {+} (2q_1 {+}
q_2)\frac{\dot{a}}{a}\frac{\dot{c}}{c} \,,
\end{equation}
\begin{equation}\label{R3}
{-} {\cal R}^{12}_{ \ \ 12} {=} (2q_1 {+} q_2)\frac{\ddot{a}}{a}
{+} q_1 \frac{\ddot{c}}{c} {+} (q_1 {+} q_2 {+}
q_3)\frac{\dot{a}^2}{a^2} {+} (2q_1 {+}
q_2)\frac{\dot{a}}{a}\frac{\dot{c}}{c} \,,
\end{equation}
\begin{equation}\label{R5}
{\cal R}^{2t}_{ \ \ 2t} = {\cal R}^{1t}_{ \ \ 1t} \,, \quad {\cal
R}^{23}_{ \ \ 23}  = {\cal R}^{13}_{ \ \ 13} \,,
\end{equation}
\begin{equation}\label{R61}
- \Re^{t}_{t} =
(2q_4+q_5)\left(2\frac{\ddot{a}}{a}+\frac{\ddot{c}}{c} \right)+
2q_4 \left(\frac{\dot{a}^2}{a^2}+2
\frac{\dot{a}}{a}\frac{\dot{c}}{c} \right) \,,
\end{equation}
\begin{equation}\label{R71}
- \Re^{1}_{1} = - \Re^{2}_{2} =
(4q_4+q_5)\left(\frac{\ddot{a}}{a}+\frac{\dot{a}}{a}\frac{\dot{c}}{c}\right)
+2q_4 \frac{\ddot{c}}{c} + (2q_4+q_5)\frac{\dot{a}^2}{a^2} \,,
\end{equation}
\begin{equation}\label{R81}
- \Re^{3}_{3} = 2q_4
\left(2\frac{\ddot{a}}{a}+\frac{\dot{a}^2}{a^2}\right) +
(2q_4+q_5)\left(\frac{\ddot{c}}{c} +
2\frac{\dot{a}}{a}\frac{\dot{c}}{c} \right) \,.
\end{equation}
Thus, we are ready to write the effective metrics in terms of
permittivity (\ref{Bepsilon}) and impermeability (\ref{Bmu}).

\subsubsection{Color (optical) metrics}

The A-metric and B-metric are, respectively,
\begin{equation}\label{Bg1}
g^{ik (A)} = \delta^{i}_{t}\delta^{k}_{t} -
\frac{1}{\varepsilon_{||} \mu_{\bot} a^2} \left(
\delta^{i}_{1}\delta^{k}_{1} + \delta^{i}_{2}\delta^{k}_{2}\right)
- \frac{1}{\varepsilon_{\bot} \mu_{\bot}
c^2}\delta^{i}_{3}\delta^{k}_{3} \,,
\end{equation}
\begin{equation}\label{Bg2}
g^{ik (B)} = \delta^{i}_{t}\delta^{k}_{t} -
\frac{1}{\varepsilon_{\bot} \mu_{||} a^2} \left(
\delta^{i}_{1}\delta^{k}_{1} + \delta^{i}_{2}\delta^{k}_{2}\right)
- \frac{1}{\varepsilon_{\bot} \mu_{\bot}
c^2}\delta^{i}_{3}\delta^{k}_{3}  \,,
\end{equation}
where $\varepsilon_{||}$, $\varepsilon_{\bot}$, $\mu_{||}$ and
$\mu_{\bot}$ are given by (\ref{Bepsilon}) and (\ref{Bmu}) with the
susceptibility tensors (\ref{R1})-(\ref{R81}). The relative
anisotropy of the permittivity/impermeability can be estimated by
the following quantities
\begin{equation}\label{ani1}
\varepsilon_{\bot} - \varepsilon_{||} = (q_2+2q_3)\left(
\frac{\ddot{c}}{c} - \frac{\ddot{a}}{a} \right) + q_2 \
\frac{\dot{a}}{a}\left( \frac{\dot{c}}{c} - \frac{\dot{a}}{a}
\right) \,,
\end{equation}
\begin{equation}\label{ani2}
\frac{1}{\mu_{\bot}} - \frac{1}{\mu_{||}} = q_2\left(
\frac{\ddot{c}}{c} - \frac{\ddot{a}}{a} \right) + (2q_3+q_2)
\frac{\dot{a}}{a}\left( \frac{\dot{c}}{c} - \frac{\dot{a}}{a}
\right) \,.
\end{equation}
It is worth noting that $q_1$ does not enter the formulas
(\ref{ani1}) and (\ref{ani2}). The isotropization of the Universe
leads to the vanishing of these differences.

\vspace{3mm}

\noindent
{\it Special case $q_2+q_3=0$}

\noindent It is clear from (\ref{R1})-(\ref{R5}) that in this
special case four non-vanishing components of the non-minimal
susceptibility coincide:
\begin{equation}\label{R6}
{\cal R}^{2t}_{ \ \ 2t} = {\cal R}^{1t}_{ \ \ 1t} = {\cal
R}^{23}_{ \ \ 23}  = {\cal R}^{13}_{ \ \ 13} \,,
\end{equation}
and, thus, the permittivity scalars are linked by
\begin{equation}\label{R7}
 \varepsilon_{\bot} \mu_{\bot} = 1 \,, \quad \varepsilon_{||} \mu_{||} =
 1\,,
\end{equation}
which provide the relations
\begin{equation}\label{om33}
{\cal V}^{(A)}_{||} = {\cal V}^{(B)}_{||} = 1  \,, \quad {\cal
V}^{(A)}_{\bot} {\cal V}^{(B)}_{\bot} = 1 \,.
\end{equation}
The color metrics in this case can also be rewritten as the
functions of only one parameter, say, $\nu^2 =
\frac{\varepsilon_{||}}{\varepsilon_{\bot}}$:
\begin{equation}\label{Bg101}
g^{ik (A)} = U^iU^k + \frac{1}{\nu^2} \Delta^{ik} +
\left(\frac{1}{\nu^2} -1 \right) X^{i}_{(3)} X^{k}_{(3)} \,,
\end{equation}
\begin{equation}\label{Bg222}
g^{ik (B)} =U^iU^k + \nu^2 \Delta^{ik} + \left(\nu^2 -1 \right)
X^{i}_{(3)} X^{k}_{(3)} \,.
\end{equation}
The structure of these color (optical) metrics shows that the
waves propagate in the longitudinal direction with the speed of
light in standard vacuum. As for the waves propagating in the
orthogonal direction, the phase velocity of one of them is less
than speed of light in vacuum, the second wave being superluminal.
We deal with birefringence.

\subsubsection{Color-acoustic metric}

Using (\ref{R61})-(\ref{R81}) one can write the color-acoustic
metric in the form
\begin{equation}
\tilde{g}^{ik} = {\cal A}^2(t) \left\{\delta^{i}_{t}\delta^{k}_{t}
- \frac{1}{a^2(t) \sigma^2_{\bot}(t)}
\left(\delta^{i}_{1}\delta^{k}_{1} +
\delta^{i}_{2}\delta^{k}_{2}\right) - \frac{1}{c^2(t)
\sigma^2_{||}(t)}\delta^{i}_{3}\delta^{k}_{3} \right\} \,,
\label{acu1}
\end{equation}
where
\begin{equation}
{\cal A}^2(t) \equiv 1-(2q_4+q_5)\left( 2\frac{\ddot{a}}{a} +
\frac{\ddot{c}}{c}\right) -2q_4 \left[\left(
\frac{\dot{a}}{a}\right)^2 + 2 \frac{\dot{a}}{a}
\frac{\dot{c}}{c}\right]
 \,, \label{acu2}
\end{equation}
\begin{equation}
\sigma^2_{||}(t) \equiv \frac{1-(2q_4+q_5)\left(
2\frac{\ddot{a}}{a} + \frac{\ddot{c}}{c}\right) -2q_4 \left[\left(
\frac{\dot{a}}{a}\right)^2 + 2 \frac{\dot{a}}{a}
\frac{\dot{c}}{c}\right]}{1-2q_4 \left[ 2\frac{\ddot{a}}{a}+
\left(\frac{\dot{a}}{a}\right)^2 \right] - (2q_4+q_5) \left(
\frac{\ddot{c}}{c} + 2 \frac{\dot{a}}{a}
\frac{\dot{c}}{c}\right)}\,, \label{acu3}
\end{equation}
\begin{equation}
\sigma^2_{\bot}(t)  \equiv \frac{1-(2q_4+q_5)\left(
2\frac{\ddot{a}}{a} + \frac{\ddot{c}}{c}\right) -2q_4 \left[\left(
\frac{\dot{a}}{a}\right)^2 + 2 \frac{\dot{a}}{a}
\frac{\dot{c}}{c}\right]}{1-(4q_4+q_5)\left(
\frac{\ddot{a}}{a}+\frac{\dot{a}}{a} \frac{\dot{c}}{c} \right) -
2q_4 \frac{\ddot{c}}{c} -
(2q_4+q_5)\left(\frac{\dot{a}}{a}\right)^2}
 \,. \label{acu4}
\end{equation}
The eikonal equation for the scalar fields reads
\begin{equation}
\tilde{g}^{ik} P_i P_k  = m^2 \,. \label{m1}
\end{equation}
Here the following definitions are introduced:
\begin{equation}
 {\cal E}^2 = m^2 {\cal A}^{-2} + P^2_{||} \sigma^{-2}_{||} +
P^2_{\bot}\sigma^{-2}_{\bot}   \,, \label{m2}
\end{equation}
\begin{equation}
 {\cal E}^2 \equiv P_t P^t \,, \quad P^2_{\bot} \equiv - (P_1
 P^1+P_2P^2)\,, \quad P^2_{||} \equiv - P_3 P^3 \,, \label{m3}
\end{equation}
where $P_i$ is particle four-momentum, ${\cal E}$ is its energy.
When a scalar particle moves along longitudinal direction, i.e.,
$P_{\bot}=0$, its three-velocity is
\begin{equation}
 v_{||}(t) = \frac{P_{||}}{{\cal E}} = \sigma_{||} \sqrt{1 - m^2 {\cal A}^{-2} {\cal E}^{-2}}
\,,\label{m4}
\end{equation}
thus, for high-energy scalar particle (${\cal E}\gg m$) the
quantity $\sigma_{||}$ gives asymptotic longitudinal velocity.
When $P_{||}=0$, one obtains, respectively,
\begin{equation}
v_{\bot}(t) = \frac{P_{\bot}}{{\cal E}} = \sigma_{\bot} \sqrt{1 -
m^2 {\cal A}^{-2} {\cal E}^{-2}} \,,\label{m41}
\end{equation}
thus,  $\sigma_{\bot}$ is some asymptotic transversal velocity.
The ratio
\begin{equation}
\frac{v_{||}(t)}{v_{\bot}(t)}=
\frac{\sigma_{||}(t)}{\sigma_{\bot}(t)}= \sqrt{
\frac{1-(4q_4+q_5)\left( \frac{\ddot{a}}{a}+\frac{\dot{a}}{a}
\frac{\dot{c}}{c} \right) - 2q_4 \frac{\ddot{c}}{c} -
(2q_4+q_5)\left(\frac{\dot{a}}{a}\right)^2}{1-2q_4 \left[
2\frac{\ddot{a}}{a}+ \left(\frac{\dot{a}}{a}\right)^2 \right] -
(2q_4+q_5) \left( \frac{\ddot{c}}{c} + 2 \frac{\dot{a}}{a}
\frac{\dot{c}}{c}\right)}}
\label{m5}
\end{equation}
does not depend on the particle energy ${\cal E}$ and is
predetermined by the values of the functions $a(t)$ and $c(t)$ and
their derivatives, as well as by the values of the non-minimal
coupling parameters $q_4$ and $q_5$. Note, that when $q_5=0$ and
the derivative coupling is absent, this ratio is equal to one.

\subsubsection{Exactly integrable example}

Let us extract from \cite{BZ05} one of the exact solutions of the
non-minimally extended Bianchi-I model with magnetic field (${\bf\Phi}=0$). This
solution is characterized by the following features. The magnetic
field is $B(t)= B(t_0)\frac{a^2(t_0)}{a^2(t)}$. It can be replaced
in our model by the ``parallel'' Yang-Mills field of the magnetic
type with $F_{12}^{(a)} = B(t_0)a^2(t_0)\delta^{(a)}_{(3)}$. The
longitudinal ${\cal P}_{||}$ and transverse ${\cal P}_{\bot}$
pressures of the matter are connected with the energy density $W$
as follows:  ${\cal P}_{||}=-{\cal P}_{\bot}= -W$. The solution
for the $a(t)$ is of de Sitter type
\begin{equation}
a(t)= a(t_0) e^{H(t-t_0)} \,, \quad \Lambda= 3H^2 \,, \label{mn0}
\end{equation}
with the constant $H$ given by
\begin{equation}
H^2 = \frac{2W(t_0)+B^2(t_0)}{10q_1 B^2(t_0) } \,, \label{mn}
\end{equation}
the coupling constants being linked by the relation
$6q_1+4q_2+q_3=0$. The solution for $c(t)$ is
\begin{equation}
c(t)= c(t_0) e^{H(t-t_0)} \frac{\left[1- \alpha
e^{-4H(t-t_0)}\right]}{(1-\alpha)} \,, \quad \alpha \equiv \kappa
q_1 B^2(t_0)\,. \label{mn1}
\end{equation}
Clearly, this non-minimal cosmological model is non-singular for
$t\geq t_0$, if $\alpha<1$, i.e., the first coupling parameter
$q_1$ satisfies inequality $0<q_1< \frac{1}{\kappa B^2(t_0)}$. For
such model the relative anisotropy can be characterized by
\begin{equation}\label{ani11}
\varepsilon_{\bot} - \varepsilon_{||} = \frac{2\kappa
(q_2+4q_3)[2W(t_0)+B^2(t_0)]}{5\left[\alpha-e^{4H(t-t_0)} \right]}
\,,
\end{equation}
\begin{equation}\label{ani22}
\frac{1}{\mu_{||}} - \frac{1}{\mu_{\bot}} = \frac{2\kappa
(q_2-2q_3)[2W(t_0)+B^2(t_0)]}{5\left[\alpha-e^{4H(t-t_0)} \right]}
\,.
\end{equation}
The anisotropy of the permittivity disappears exponentially at $t\to
\infty$, i.e., when Bianchi-I model isotropizes. Mention that
$\varepsilon_{\bot} = \varepsilon_{||}$, when $q_2+4q_3=0$, as well
as, $\frac{1}{\mu_{||}} = \frac{1}{\mu_{\bot}}$, when $q_2=2q_3$,
nevertheless, it can not occur simultaneously, since when
$q_2=q_3=0$, $q_1$ also vanishes and the non-minimal model
degenerates. Analogously, the color-acoustic anisotropy is
characterized by the relation
\begin{equation}\label{ani23}
\left(\frac{\sigma_{||}}{\sigma_{\bot}}\right)^2 = \frac{1 -
\alpha e^{-4H(t-t_0)} - H^2 \left[3(4q_4+q_5) + (q_5-12q_4)\alpha
e^{-4H(t-t_0)} \right]}{[1-3H^2(4q_4+q_5)]\left[1 - \alpha
e^{-4H(t-t_0)} \right]}   \,.
\end{equation}
Of course, at $t\to \infty$, this quantity tends to one,
$\frac{\sigma_{||}}{\sigma_{\bot}} \to 1$, as it should be.

The analysis of the expressions (\ref{om3}) for the given model
shows, that for generic $q_2$ and $q_3$ there are, in principle,
two critical moments of time, when the phase velocities vanish and
the A-wave or/and B-wave stop. At the first moment, $t^{*}_1$, the
phase velocities of both waves propagating in the longitudinal
direction, as well as of the A-wave moving in the orthogonal
direction, vanish. It is possible, when
\begin{equation}\label{ve1}
1+2{\cal R}^{23}_{\ \ 23}(t^*_1) =0  \quad \rightarrow \quad t^*_1
= t_0 + \frac{1}{4H} \ln{\left\{\frac{\alpha [1+2H^2
(3q_2+4q_3)]}{1+2H^2 q_2}\right\}}\,,
\end{equation}
if the argument of logarithm is more than one. At the second
critical moment, $t^{*}_2$, the B-wave propagating in the
transverse directions has vanishing phase velocity, it is
possible, when
\begin{equation}\label{ve2}
1+2{\cal R}^{12}_{\ \ 12}(t^*_2) =0  \quad \rightarrow \quad t^*_2
= t_0 + \frac{1}{4H} \ln{\left\{\frac{\alpha [1+10H^2q_2)]}{1+2H^2
q_2}\right\}}\,,
\end{equation}
if the argument of this logarithm is also more than one. In its
turn, we can choose the parameters $\alpha$, $q_2$ and $q_3$ so
that both optical metrics are regular at each time. Let us
illustrate this possibility by the example, when $q_2=-2q_1<0$ and
$q_3= 2q_1>0$, satisfying the basic condition $6q_1+4q_2+q_3=0$.
For this model the waves moving in the longitudinal direction have
the phase velocity equal to one, as for the waves propagating in
the transverse direction, they are characterized by the following
reduced formulas
\begin{equation}\label{arc1}
\left({\cal V}^{(A)}_{\bot}\right)^2 = \frac{(1-4q_1 H^2) - \alpha
(1+4q_1 H^2)e^{-4H(t-t_0)}}{(1-4q_1 H^2) - \alpha (1-20q_1
H^2)e^{-4H(t-t_0)}} \,, \quad {\cal V}^{(B)}_{\bot} {\cal
V}^{(A)}_{\bot} = 1 \,.
\end{equation}
Taking (\ref{mn}) into account, we obtain that the square of the
phase velocity (\ref{arc1}) is positive for $t>t_0$, when
\begin{equation}\label{arc10}
\frac{2}{5} < \alpha < \frac{3}{7} \,, \quad \frac{4
W(t_0)}{B^2(t_0)} < \min \left\{ \frac{5\alpha-2}{6-5\alpha} \,,
\frac{3-7\alpha}{1+\alpha}\right\} \,.
\end{equation}
Analogously, the longitudinal and transversal color-acoustic waves
stop, when, respectively,
\begin{equation}\label{arc2}
t^*_3 = t_0 + \frac{1}{4H} \ln{\left\{\frac{\alpha
[1+H^2(q_5-12q_4)]}{1-3H^2(4q_4+q_5)}\right\}}\,,
\end{equation}
\begin{equation}\label{arc3}
t^*_4 = t_0 + \frac{1}{4H} \ln{\alpha}\,.
\end{equation}
The color-acoustic metric degenerates when ${\cal A}(t^*_5)=0$,
where
\begin{equation}\label{arc4}
t^*_5 = t_0 + \frac{1}{4H}
\ln{\left\{\frac{\alpha[1-H^2(12q_4+11q_5)]}{1-3H^2(4q_4+q_5)}\right\}}\,.
\end{equation}
Let us mention that the color-acoustic metrics does not exist initially, if
the non-minimal parameters are coupled by
\begin{equation}\label{arc5}
10q_1 = (4q_4+q_5)[2W(t_0)+B^2(t_0)] \,.
\end{equation}
To avoid the singularities in the color-acoustic metric one can put,
e.g., $q_5=-4q_4<0$, providing the expressions (\ref{acu2}) and
(\ref{ani23}) to be positive for $t>t_0$.

\subsection{Spatially flat FLRW model}

FLRW model with $k=0$ (i.e., spatially flat model) can be obtained
from the previous model, if we put $c(t)=a(t)$. The formulas for the
A- and B- metrics can be converted into
\begin{equation}\label{F1}
g^{ik (A)} = g^{ik (B)} = \delta^{i}_{t}\delta^{k}_{t} -
\frac{1}{\varepsilon \mu a^2} \left( \delta^{i}_{1}\delta^{k}_{1}
+ \delta^{i}_{2}\delta^{k}_{2} + \delta^{i}_{3}\delta^{k}_{3}
\right) \,,
\end{equation}
where
\begin{equation}\label{F3}
\varepsilon \equiv 1 -2 (3q_1+2q_2+q_3)\frac{\ddot{a}}{a} -2
(3q_1+q_2)\left(\frac{\dot{a}}{a}\right)^2 \,,
\end{equation}
\begin{equation}\label{F4}
\mu \equiv 1 -2 (3q_1+q_2)\frac{\ddot{a}}{a} -2
(3q_1+2q_2+q_3)\left(\frac{\dot{a}}{a}\right)^2  \,.
\end{equation}
Effective refractive index $n_{({\rm eff})}$ is given by
\begin{equation}\label{F5}
n^2_{({\rm eff})} \equiv \varepsilon \mu a^2 = a^2 \ \frac{\left[a^2 -2
(3q_1+2q_2+q_3)\ddot{a} a -2 (3q_1+q_2) {\dot{a}}^2
\right]}{\left[ a^2 -2 (3q_1+q_2)\ddot{a}a -2
(3q_1+2q_2+q_3){\dot{a}}^2 \right]} \,.
\end{equation}
The effective refractive index coincides with the standard one,
i.e., $n_{({\rm eff})}=a(t)$,  first, when $q_2+q_3=0$, second, when
$\frac{\dot{a}}{a}=H_0=\hbox{const}$. When the law of isotropic
expansion is the power-like one, say, $a(t)= a(t_0)
\left(\frac{t}{t_0}\right)^{\omega}$, then (\ref{F5}) reduces to
\begin{equation}
\frac{n^2_{({\rm eff})}}{a^2(t)} = \frac{t^2 -
2\omega\left[\omega(6q_1+3q_2+q_3)-(3q_1+2q_2+q_3) \right]}{t^2 -
2\omega\left[\omega(6q_1+3q_2+q_3)-(3q_1+q_2) \right]} \,.
\label{F121}
\end{equation}
At the moment $t^*$
\begin{equation}
t^* = \sqrt{2\omega\left[\omega(6q_1+3q_2+q_3)-(3q_1+q_2) \right]}
\label{F122}
\end{equation}
the color (optical) wave stops, since the phase velocity $V_{ph}=
\frac{a}{n_{({\rm eff})}}$ vanishes, of course, the expression in the square root
has to be positive. There is a number of sub-models in which the
singularities do not appear. For instance, when $q_1=-q$, $q_2=2q$,
$q_3=-q$, $q>0$ (this model has been considered in \cite{MH,BL05}
for the spherically symmetric static case) and $\omega>1$, the
expression (\ref{F121}) is positive for each time. Color-acoustic
metric reads
\begin{equation}
\tilde{g}^{ik} = \tilde{{\cal
A}}^2(t)\left[\delta^{i}_{t}\delta^{k}_{t} -
\frac{1}{a^2(t)\sigma^2(t)} \left( \delta^{i}_{1}\delta^{k}_{1} +
\delta^{i}_{2}\delta^{k}_{2} +
\delta^{i}_{3}\delta^{k}_{3}\right)\right] \,. \label{F12}
\end{equation}
\begin{equation}
\tilde{{\cal A}}^2(t) \equiv 1-3(2q_4+q_5)\frac{\ddot{a}}{a} -6q_4
\left(\frac{\dot{a}}{a}\right)^2
 \,, \label{F56}
\end{equation}
\begin{equation}
\sigma^2(t) = \sigma^2_{\bot}(t) = \sigma^2_{||}(t) =
 \frac{1-3(2q_4+q_5)\frac{\ddot{a}}{a} -6 q_4 \left(
\frac{\dot{a}}{a}\right)^2}{1-(6q_4+q_5) \frac{\ddot{a}}{a} -
(6q_4+2q_5)\left(\frac{\dot{a}}{a}\right)^2}\,, \label{F7}
\end{equation}
The quantity $\sigma$ differs from one if and only if $q_5 \neq 0$,
i.e., the derivative coupling of the Higgs fields with curvature is
present. Finally, this color-acoustic metric is non-singular, if,
for instance, $q_5=-2q_4>0$ and $\omega>\frac{2}{3}$.

\subsection{De Sitter model}

We consider now the model with positive curvature, $K>0$, and
reduce the metric to the de Sitter form \cite{Weinberg}
\begin{equation}\label{deSitter}
ds^2 = dt^2 - \exp\{2 \sqrt{K} t\} (\delta_{\alpha \beta}
dx^{\alpha} dx^{\beta}) \,.
\end{equation}
For this spacetime the Riemann, Ricci tensors and Ricci scalar,
respectively, take the form
\begin{equation}\label{curv}
R_{ikmn}=-K\left(g_{im}g_{kn}-g_{in}g_{km}\right) \,, \quad
R_{ik}=-3Kg_{ik}\,, \quad R=-12K \,.
\end{equation}
The tensors ${\cal R}_{ikmn}$ (\ref{sus}) and $\Re_{ik}$
(\ref{Reqq}) yield
\begin{equation}\label{simpa}
{\cal
R}_{ikmn}=-K(6q_1+3q_2+q_3)\left(g_{im}g_{kn}-g_{in}g_{km}\right)\,,\quad
\Re_{ik}=-3K(4q_4+q_5)g_{ik}\,.
\end{equation}
Let us mention that, when we deal with de Sitter model, only two combinations of the
coupling parameters $\lambda_1 {=} 6q_1{+}3q_2{+}q_3$ and
$\lambda_4 {=} q_4 {+} \frac{1}{4} \lambda_5$ (see (\ref{irredu7})) enter
the susceptibility tensors ${\cal R}_{ikmn}$ and $\Re_{ik}$. In this
subsection we use the parameters $\lambda_1$ and $\lambda_4$ instead of $q_1$,...$q_5$
in order to shorten the formulas.
Then, the ${\bf H}^{ik}$ tensor and the ${\bf \Psi}^m$ vector
simplify significantly, and the equations (\ref{33YMeq}) and
(\ref{Heq}) convert, respectively, into
\begin{equation}\label{1simpa}
 \left(1-2\lambda_1 K \right)  \D_k {\bf F}^{ik} =
{\cal G}^2 \left(1-12 \lambda_4 K\right) [\D^i {\bf \Phi},{\bf \Phi}] \,,
\end{equation}
\begin{equation}\label{2simpa}
\left(1-12 \lambda_4 K \right) \D_k \D^k {\bf \Phi} =  - m^2 {\bf\Phi} \,.
\end{equation}
Thus, when $\lambda_1< \frac{1}{2K}$  and
$\lambda_4<\frac{1}{12K}$ the color and color-acoustic metrics are
simply proportional to the spacetime one
\begin{equation}\label{3simpa}
g^{(A)ik}= g^{(B)ik} = g^{*ik}= \sqrt{1-2 \lambda_1 K} \
g^{ik}\,,
\end{equation}
\begin{equation}\label{4simpa}
\tilde{g}^{(A)ik}= \tilde{g}^{(B)ik} = \sqrt{1-12 \lambda_4 K} \
g^{ik}\,.
\end{equation}
One can redefine the coupling constant ${\cal G}$ and the mass $m$
\begin{equation}\label{5simpa}
{\cal G} \rightarrow {\cal G}^{*} = {\cal G}
\sqrt{\frac{1{-}12 \lambda_4 K}{1{-}2 \lambda_1 K}} \,,
\quad m \rightarrow m^{*} = m \sqrt{\frac{1}{1{-}12 \lambda_4 K}}
\,,
\end{equation}
so that the Yang-Mills and Higgs equations take the standard
(minimal) form, but the constants are non-minimally redefined. In
the WKB-approximation the conditions $g^{*ik}p_ip_k=0$ and
$g^{ik}p_ip_k=0$ are in fact equivalent, thus, the massless test
particles propagate in the ``non-minimally active'' de Sitter
model as in standard de Sitter one. This fact has been emphasized
above during the analysis of the FLRW model. The special case,
when $2K(6q_1{+}3q_2{+}q_3){=}1{=} 2 \lambda_1 K$ and $3K(4q_4{+}q_5){=}1{=} 12 \lambda_4 K$,
related to the so-called hidden Yang-Mills and Higgs fields, is discussed in
\cite{BDZ}.

\section{Conclusions}

\noindent 1. We developed the concept of effective metrics and
introduced the associated, color and color-acoustic metrics, attributed to the
Einstein-Yang-Mills-Higgs model. The procedure of reconstruction of the effective
metrics consists of two steps. The first one, the reconstruction of the associated
metrics, is based on an original multi-metric representations (\ref{supergeneral}) and
(\ref{aca1}) of the constitutive tensors for the Yang-Mills and Higgs fields, respectively.
The second step, the identification of the associated metrics with color
and color-acoustic ones, is provided by the WKB-analysis of the
master equations and by the well-known analogy with optical metrics.

\noindent 2. Curvature interactions between gravitational, gauge and
scalar fields are described in terms of five-parameter non-minimal
EYMH model, and the basic ideas concerning associated, color and
color-acoustic metrics are realized. The color (see (\ref{Bg1u}), (\ref{Bg2u})
with (\ref{Bepsilon}), (\ref{Bmu}))  and the color-acoustic (\ref{acumet11})
metrics for the uniaxial models are reconstructed explicitly.

\noindent 3. The formalism of effective metrics is applied to
Bianchi-I, FLRW and de Sitter cosmological models, and the
associated, color and color-acoustic metrics are represented
explicitly in terms of metric coefficients of the corresponding
spacetimes and their derivatives. Phase velocities of the
corresponding non-minimally coupled color (\ref{om3}) and
color-acoustic (\ref{acumet111}) waves are the main distinguishing
features in these expressions. These phase velocities and their
ratios are calculated for all models (see, e.g., (\ref{acu3}),
(\ref{acu4}), (\ref{m5}), (\ref{ani23}), (\ref{arc1})). Thus, we
obtained the tool for a qualitative analysis of the problem of
propagation of test color, optical and scalar waves in the
``non-minimally active'' vacuum, interacting with curvature.

\noindent 4. The analysis of the presented models allows us to
conclude that for generic set of coupling constants $q_1$, $q_2$,
\dots\ $q_5$ the color and color-acoustic metrics are singular.
This means that generally the phase velocities of the
corresponding waves can be equal to zero or infinity at specific
time moments (see, e.g., (\ref{ve1}), (\ref{ve2}), (\ref{arc2}), (\ref{F122})).
The first case relates to the stopping of the
corresponding wave. The second one is associated with the
appearance of the analogs of trapped surfaces, apparent horizons
or event horizons, which are discussed in detail in the literature
devoted to analog gravity \cite{Visser1,Novello,VolovikBook}.

\noindent 5. In its turn we present the examples of the sets of
coupling constants $q_1$, $q_2$, \dots\ $q_5$, for which there are
no singularities in color and color-acoustic metrics. Such (regular)
models predict, in particular, that the phase velocities of the
waves essentially depend on the time moment, on the direction of
propagation and are very sensitive to the choice of the values of the
free parameters of the model. These conclusions are in agreement with the
ones made in numerous papers, devoted to the variation of the speed
of light under the influence of curvature interactions in
electrodynamics (see, e.g.,
\cite{Lafrance} - \cite{Solanki} and
references therein). These conclusions require to attract  a special
attention to the analysis of the birefringence and time delay
effects in observational astronomy, since the photons registered simultaneously
are not necessarily emitted at the same time. We hope to devote a special paper to
the study of such problems.

\appendix

\section{Symmetry of the constitutive tensors}\label{appa}

\subsection{Tensor $C^{ikmn}_{(a)(b)}$}

Model under discussion (see (\ref{1act}) and (\ref{sus})) is characterized by the tensor
$C^{ikmn}_{(a)(b)}$ symmetric with respect to indices $(a)$ and $(b)$, as well as symmetric with
respect to transposition of the pairs of indices $ik$ and $mn$. In the general
case this symmetry can be broken, and the tensor
$C^{ikmn}_{(a)(b)}$ can loose the symmetry properties $C^{ikmn}_{(a)(b)} = C^{ikmn}_{(b)(a)}$ and
$C^{ikmn}_{(a)(b)} = C^{mnik}_{(a)(b)}$ separately. For $U(1)$ gauge
invariant electrodynamics the antisymmetric part of $C^{ikmn}$:
\begin{equation}
C_{({\rm skewon})}^{ikmn} \equiv \frac{1}{2}(C^{ikmn}- C^{mnik})
\,, \label{skewon1}
\end{equation}
introduced by Hehl and Obukhov \cite{HehlObukhov} and indicated as
skewon, disappears from the Lagrangian. In the $SU(n)$ symmetric
Yang-Mills-Higgs theory ($n>1$) the skewonic terms can appear in
the Lagrangian explicitly, since the strength field tensor
$F^{(a)}_{ik}$ has a group index with two and more values, and
indices $(a)$ and $(b)$ can be in the antisymmetric combination.
Generally, we have the following decomposition of the second term in the Lagrangian (\ref{0act}):
\begin{equation}
\frac{1}{2} C^{ikmn}_{(a)(b)} F^{(a)}_{ik} F^{(b)}_{mn} {=}
 \frac{1}{2}\Theta^{ikmn}_{(a)(b)} F^{(a)}_{ik} F^{(b)}_{mn} {+}
\frac{1}{2}\Xi^{ikmn}_{(a)(b)} F^{(a)}_{ik}F^{(b)}_{mn} \,, \label{csym}
\end{equation}
where the tensors $\Theta^{ikmn}_{(a)(b)}$ and $\Xi^{ikmn}_{(a)(b)}$ possess the following symmetry of indices
\begin{equation}
\Theta^{ikmn}_{(a)(b)} = \Theta^{mnik}_{(a)(b)} = \Theta^{ikmn}_{(b)(a)} \,,
\quad \Xi^{ikmn}_{(a)(b)} = - \Xi^{mnik}_{(a)(b)} = \Xi^{mnik}_{(b)(a)} \,.
\label{csym1}
\end{equation}
The term $\Xi^{ikmn}_{(a)(b)}$ explicitly represents
the Yang-Mills skewon, if we use the terminology introduced by Hehl
and Obukhov \cite{HehlObukhov}. In this paper we do not consider
such a model and assume that $\Xi^{ikmn}_{(a)(b)} =0$.

The symmetric  part of the tensor of linear response $\Theta^{ikmn}_{(a)(b)}$ can be decomposed
using the velocity four-vector $U^i$ in the same way as in electrodynamics:
$$
\Theta^{ikmn}_{(a)(b)} = \frac{1}{2} \left(
\varepsilon^{im}_{(a)(b)} U^k U^n - \varepsilon^{in}_{(a)(b)} U^k
U^m + \varepsilon^{kn}_{(a)(b)} U^i U^m -
\varepsilon^{km}_{(a)(b)} U^i U^n \right)
$$
\begin{equation}
{}+\frac12 \left[ -\eta^{ik}_{\ \ l}(\mu^{-1})^{ls}_
{(a)(b)} \eta^{mn}_{\ \ s} + \eta^{ik}_{\ \ l}(U^m
\nu^{ln}_{(a)(b)} - U^n \nu_{(a)(b)}^{l m}) + \eta^{\ mn}_l
(U^i \nu_{(a)(b)}^{l k} - U^k \nu_{(a)(b)}^{li} ) \right] \,.
\label{Cdecomp}
\end{equation}
Here $\varepsilon^{im}_{(a)(b)}$ and  $(\mu^{-1})^{pq}_{(a)(b)}$ are
analogs of the tensors of dielectric permittivity and magnetic
impermeability, respectively, and $\nu_{(a)(b)}^{p m}$ is an analog
of the tensor of magneto-electric coefficients. These quantities are
defined as
\begin{equation}
\varepsilon^{im}_{(a)(b)} = 2 \Theta^{ikmn}_{(a)(b)} U_k
U_n \,, \quad (\mu^{-1})^{pq}_{(a)(b)} = - \frac{1}{2} \eta^p_{\
ik} \Theta^{ikmn}_{(a)(b)} \eta^{ \ \ q}_{mn} \,,
\label{emu}
\end{equation}
\begin{equation}
\nu_{(a)(b)}^{p m} = \eta^p_{\ \ ik} \Theta^{ikmn}_{(a)(b)} U_n = U_k \Theta^{mkln}_{(a)(b)}
\eta_{ln }^{ \ \ p} \,. \label{nu}
\end{equation}
Tensors $\eta_{mnl}$ and $\eta^{ikl}$ are the antisymmetric
tensors orthogonal to $U^i$ defined as
\begin{equation}
\eta_{mnl} \equiv \epsilon_{mnls} U^s \,, \quad \eta^{ikl} \equiv
\epsilon^{ikls} U_s \,. \label{eta}
\end{equation}
They are connected by the useful identities
\begin{equation}
- \eta^{ikp} \eta_{mnp} = \delta^{ikl}_{mns} U_l U^s = \Delta^i_m
\Delta^k_n - \Delta^i_n \Delta^k_m \,, \quad \frac{1}{2}
\eta^{ikl}  \eta_{klm} = - \delta^{il}_{ms} U_l U^s = - \Delta^i_m
\,, \label{etaid1}
\end{equation}
where $\delta^{ikl}_{mns}$ and $\delta^{il}_{ms}$ are the
generalized Kronecker tensors \cite{MTW}. The symmetric projection
tensor $\Delta^{ik}$ is defined as $\Delta^{ik} = g^{ik} - U^i U^k
$. The tensors $\varepsilon^{ik}_{(a)(b)}$ and $(\mu^{-1})^{ik}_{
(a)(b)}$ are symmetric, but $\nu^{lk}_{(a)(b)}$ is generally
non-symmetric with respect to the spacetime indices. These three
tensors are orthogonal to $U^i$,
\begin{equation}
\varepsilon^{ik}_{(a)(b)} U_k = 0, \quad (\mu^{-1})^{ik}_{(a)(b)}
U_k = 0, \quad \nu_{(a)(b)}^{l k} U_l = 0 = \nu_{(a)(b)}^{lk} U_k
\,, \label{orto}
\end{equation}
and possess 21 independent components for any fixed set of the
indices $(a)$ and $(b)$.

To complete the analogy, we introduce two four-vectors of
excitation ${\cal D}^i_{(a)}$ and ${\cal H}^i_{(a)}$, as well as,
two four-vectors of the field strength ${\cal E}^i_{(a)}$ and
${\cal B}^i_{(a)}$ by the definitions
\begin{equation}
{\cal D}^i_{(a)} = {\cal H}^{ik}_{(a)} U_k \,, \quad {\cal
H}^i_{(a)} = {\cal H}^{*ik}_{(a)} U_k \,, \quad {\cal E}_i^{(a)} =
F^{(a)}_{ik} U^k \,, \quad {\cal B}_i^{(a)} = F^{*(a)}_{ik} U^k
\,, \label{DHEB}
\end{equation}
where the four-vector of the velocity $U^i$ is normalized by
unity: $U^iU_i=1$. These vectors are orthogonal to the velocity
four-vector $U^i$:
\begin{equation}
{\cal D}^i_{(a)} U_i = 0 = {\cal E}_i^{(a)} U^i \,, \quad {\cal
H}^i_{(a)} U_i = 0 = {\cal B}_i^{(a)} U^i \,, \label{orthogon}
\end{equation}
and form the basis for the decomposition of the $F^{(a)}_{mn}$ and
${\cal H}^{mn}_{(a)}$ tensors:
\begin{equation}
F^{(a)}_{mn} = {\cal E}^{(a)}_m U_n - {\cal E}^{(a)}_n U_m -
\epsilon_{mn \cdot \cdot}^{ \ \ \ ls}U_s {\cal B}_l^{(a)} \,,
\quad {\cal H}^{mn}_{(a)} = {\cal D}^m_{(a)} U^n - {\cal
D}^n_{(a)} U^m - \epsilon^{mn}_{\cdot \cdot \ ls}U^s {\cal
H}^l_{(a)} \,. \label{FHdecomp}
\end{equation}
When we deal with the five-parameter EYMH model, the direct
calculations using (\ref{HikR}) show that
\begin{equation}
\varepsilon^{im}_{(a)(b)} = \left[\Delta^{im} + 2 {\cal R}^{ikmn}
U_k U_n \right] G_{(a)(b)} \,, \label{Re}
\end{equation}
\begin{equation}
(\mu^{-1})^{pq}_{(a)(b)} = G_{(a)(b)} \left[ \Delta^{pq}- 2 \
^{*}{\cal R}^{*plqs} U_l U_s \right]  \,, \label{Rmu}
\end{equation}
\begin{equation}
\nu^{pm}_{(a)(b)} =  - ^{*}{\cal R}^{plnm} U_l U_n  G_{(a)(b)} \,.
\label{Rnu}
\end{equation}
Thus, one can state that the non-minimal coupling of the
gravitational and Yang-Mills-Higgs field effectively changes the
properties of the vacuum linear response. Particularly, the tensor
of non-minimal susceptibility ${\cal R}^{ijkl}$ determines the
variation of color electric-type permittivity, the corresponding
double dual tensor $^{*}{\cal R}^{*plqs}$ is responsible for the
changes in the color magnetic-type impermeability tensor, and
non-vanishing $^{*}{\cal R}^{plqs}$ produces the so-called color
magneto-electric-type properties of the non-minimal vacuum.

\subsection{Tensor ${\cal C}^{ik}_{(a)(b)}$}

The third term in the Lagrangian (\ref{0act}) also contains two parts
\begin{equation}
-{\cal C}^{ik}_{(a)(b)} \hat{D}_i \Phi^{(a)} \hat{D}_k \Phi^{(b)} = -\Omega^{ik}_{(a)(b)} \hat{D}_i
\Phi^{(a)} \hat{D}_k \Phi^{(b)}
 - \Psi^{ik}_{(a)(b)} \hat{D}_i  \Phi^{(a)} \hat{D}_k \Phi^{(b)} \,, \label{C2}
\end{equation}
where tensors $\Omega^{ik}_{(a)(b)}$ and $\Psi^{ik}_{(a)(b)}$ are
symmetric and antisymmetric, respectively, i.e.,
\begin{equation}
\Omega^{ik}_{(a)(b)} = \Omega^{ki}_{(a)(b)} = \Omega^{ik}_{(b)(a)} \,, \quad
\Psi^{ik}_{(a)(b)} = - \Psi^{ki}_{(a)(b)} = \Psi^{ki}_{(b)(a)} \,. \label{antisym}
\end{equation}
The decomposition of the symmetric tensor $\Omega^{ik}_{(a)(b)}$ has the same structure as a symmetric
stress-energy tensor
\begin{equation}
\Omega^{ik}_{(a)(b)} =  {\cal A}_{(a)(b)} U^i U^k  + {\cal
J}^{i}_{(a)(b)} U^k + {\cal J}^{k}_{(a)(b)} U^i + {\cal
B}^{ik}_{(a)(b)}\,, \label{acou1}
\end{equation}
where ${\cal J}^{i}_{(a)(b)}$ and ${\cal B}^{ik}_{(a)(b)}$ are
orthogonal to the velocity four-vector $U^k$. For the non-minimal
vacuum we obtain
\begin{equation}
{\cal A}_{(a)(b)} =  G_{(a)(b)} \left(1+ q_4 R + q_5 R^{ik} U_iU_k
\right) \,, \label{acou2}
\end{equation}
\begin{equation}
{\cal J}^{i}_{(a)(b)} = G_{(a)(b)} q_5 \left( R^{im} - U^i
R^{mn} U_n \right) U_m \,, \label{acou3}
\end{equation}
\begin{equation}
\quad {\cal B}^{ik}_{(a)(b)}= G_{(a)(b)} \left[ (1+q_4R)
\Delta^{ik} + q_5 R^{mn}\Delta^{i}_m \Delta^{k}_n \right]\,.
\label{acou4}
\end{equation}
The skew-symmetric part $\Psi^{ik}_{(a)(b)}$ has the
decomposition analogous to the Maxwell tensor
\begin{equation}
\Psi^{ik}_{(a)(b)} = {\cal M}^i_{(a)(b)} U^k -
{\cal M}^k_{(a)(b)} U^i - \eta^{ikl} \Lambda_{l(a)(b)} \,,
\label{acou6}
\end{equation}
with
\begin{equation}
{\cal M}^i_{(a)(b)} = - {\cal M}^i_{(b)(a)} \,, \quad
\Lambda_{l(a)(b)} = - \Lambda_{l(b)(a)} \,,
\label{acou61}
\end{equation}
and introduces the scalar skewons by the analogy
with the terminology of Hehl and Obukhov.

\subsection{Tensor ${\cal C}_{ijkl}^{\ \ \ \ mnpq}$}

This tensor can also be decomposed into irreducible parts using
the same procedure as in case of $C^{ijkl}_{(a)(b)}$,
nevertheless, we do not present this procedure here.

\section{WKB-approximation for the gauge field in
the uniaxial case}\label{appb}

In the WKB-approximation the gauge potentials $A_k^{(a)}$ and the
field strengths $F_{kl}^{(a)}$ can be extrapolated as follows
\begin{equation}
A_k^{(a)} \rightarrow {\cal A}_k^{(a)} e^{i \Psi} \,, \quad
F_{kl}^{(a)} \rightarrow i \left[ p_k {\cal A}_l^{(a)} - p_l {\cal
A}_k^{(a)}\right] e^{i \Psi} \,, \quad p_k = \nabla_k \Psi
\,.\label{AP1}
\end{equation}
Let us mention that the nonlinear terms in (\ref{Fmn}) give the values of
the next order in WKB-approximation, thus, such a model of gauge
field is effectively Abelian. In the leading order approximation
the Yang-Mills equations are reduced to
\begin{equation}
C^{ikmn}_{(a)(b)} \ p_k \ p_m \ {\cal A}_n^{(b)} = 0 \,.
\label{AP2}
\end{equation}
Substitution of $C^{ikmn}_{(a)(b)}$ from (\ref{maindecomp}) with
$g^{ik(A)}$ from (\ref{Bg1u}) and $g^{ik(B)}$ from (\ref{Bg2u})
with $\hat{\mu}$ and $\gamma$ given by (\ref{Bg111}) yields
$$
 g^{im (A)} p_m  \left[g^{kn(A)}p_k {\cal A}_n^{(a)} \right] -
 g^{in (A)} {\cal A}_n^{(a)} \left[g^{km(A)}p_k p_m \right] =
$$
\begin{equation} \label{AP3}
= \frac{1}{\varepsilon^2_{||}\mu^2_{\bot}} \left(1-
\frac{\varepsilon_{||}\mu_{\bot}}{\varepsilon_{\bot}\mu_{||}}\right)
p_k p_m  {\cal A}_n^{(a)} \left[ \left(X^{i}_{(1)}X^{k}_{(2)} -
X^{i}_{(2)}X^{k}_{(1)} \right) \left(X^{m}_{(1)}X^{n}_{(2)} -
X^{m}_{(2)}X^{n}_{(1)} \right)\right] \,.
\end{equation}
Projection of this equations onto the velocity four-vector $U_i$
gives the scalar ratio
\begin{equation} \label{AP4}
\left(U^k p_k \right) \left[g^{mn(A)} p_m {\cal A}_n^{(a)}\right]
= \left(U^n {\cal A}_n^{(a)}\right) \left[g^{km (A)} p_k p_m
\right]\,,
\end{equation}
which is satisfied if, for instance, we use the Landau gauge $U^n
{\cal A}_n^{(a)}=0$ and the condition of orthogonality of the wave
four-vector and amplitude four-vector in the first associated
metric, i.e., $g^{kn} p_k {\cal A}_n^{(a)}=0 $. Projections onto
the axes, given by $X^i_{(1)}$, $X^i_{(2)}$ and $X^i_{(3)}$ yield,
respectively,
$$
{\cal A}_{(1)}^{(a)} \left[ g^{km (A)} p_k p_m + p_{(2)}^2 \left(
 \frac{1}{\varepsilon_{||}\mu_{\bot}} - \frac{1}{\varepsilon_{\bot}\mu_{||}}\right) \right]-
 {\cal A}_{(2)}^{(a)} \left[ p_{(1)}p_{(2)} \left(
 \frac{1}{\varepsilon_{||}\mu_{\bot}} - \frac{1}{\varepsilon_{\bot}\mu_{||}}\right)
 \right]= 0 \,,
$$
$$
{\cal A}_{(1)}^{(a)} \left[ p_{(1)}p_{(2)} \left(
 \frac{1}{\varepsilon_{||}\mu_{\bot}} - \frac{1}{\varepsilon_{\bot}\mu_{||}}\right)
 \right] - {\cal A}_{(2)}^{(a)} \left[ g^{km (A)} p_k p_m + p_{(1)}^2 \left(
 \frac{1}{\varepsilon_{||}\mu_{\bot}} - \frac{1}{\varepsilon_{\bot}\mu_{||}}\right) \right]= 0 \,,
$$
\begin{equation} \label{AP5}
{\cal A}_{(3)}^{(a)} \left[g^{km (A)} p_k p_m \right]= 0 \,,
\end{equation}
where ${\cal A}_{(1)}^{(a)} \equiv X^k_{(1)}{\cal A}_{k}^{(a)}$,
$p_{(1)} \equiv X^k_{(1)}p_k$, etc. Taking into account the
relation
\begin{equation} \label{AP6}
\left[g^{km (B)} p_k p_m \right]= \left[g^{km (A)} p_k p_m \right]
+ \left( p_{(1)}^2 + p_{(2)}^2\right)\left(
 \frac{1}{\varepsilon_{||}\mu_{\bot}} - \frac{1}{\varepsilon_{\bot}\mu_{||}}\right)
 \,,
\end{equation}
one can conclude that nontrivial solution of (\ref{AP5}) exists,
when
\begin{equation} \label{AP7}
\left[g^{km (A)} p_k p_m \right]\left[g^{lj (B)} p_l p_j \right]=
0 \,.
\end{equation}
Thus, the associated metrics $g^{km (A)}$ and $g^{km (B)}$ are the
color ones.

\section{Alternative description of the particle motion in terms of tidal
force}\label{appc}

Particle dynamics can be alternatively described using the
equation of motion with effective force. Instead of equation of
null geodesics in the effective spacetime with metric
$g^{ik(\alpha)}$, where $\alpha = A,B$, one can write the equation
\begin{equation} \label{AP31}
\frac{d^2 x^i}{d\tau^2} + \Gamma^i_{kl} \frac{dx^k}{d\tau}
\frac{dx^l}{d\tau} = F^{i (\alpha)} \,,
\end{equation}
where
\begin{equation} \label{AP32}
 F^{i (\alpha)} \equiv \Pi^{i(\alpha)}_{kl} \frac{dx^k}{d\tau}
\frac{dx^l}{d\tau} \,, \quad \Pi^{i(\alpha)}_{kl} \equiv
\Gamma^i_{kl} - \Gamma^{i(\alpha)}_{kl} \,.
\end{equation}
$\Gamma^i_{kl}$ and $\Gamma^{i(\alpha)}_{kl}$ are the Christoffel
symbols for the real and effective spacetimes, respectively. The
quantity $\Pi^{i(\alpha)}_{kl}$, the difference of the Christoffel
symbols symmetric with respect to indices $k$ and $l$, is known to
be a tensor, thus, the quantity $F^{i (\alpha)}$ is a vector.
Since we consider the interaction of Yang-Mills and Higgs field
with spacetime curvature, we can indicate this force as a tidal
one. The tidal force $F^{i (\alpha)}$ is quadratic in the particle
velocity four-vector and is predetermined by the structure of the tensor
$\Pi^{i(\alpha)}_{kl}$. For Bianchi-I model this tensor has
the following non-vanishing components:
$$
\Pi^{t(A)}_{11} = \Pi^{t(A)}_{22} = \frac{1}{2} \frac{d}{dt}\left[
a^2\left(1{-}\varepsilon_{||}\mu_{\bot} \right) \right] \,, \quad
\Pi^{t(B)}_{11} = \Pi^{t(B)}_{22} = \frac{1}{2} \frac{d}{dt}\left[
a^2\left(1{-}\varepsilon_{\bot}\mu_{||} \right) \right] \,,
$$
$$
\Pi^{t(A)}_{33} = \Pi^{t(B)}_{33} = \frac{1}{2} \frac{d}{dt}\left[
c^2\left(1{-}\varepsilon_{\bot}\mu_{\bot} \right) \right] \,,
\quad \Pi^{1(A)}_{1t}= \Pi^{2(A)}_{2t} = - \frac{1}{2}
\frac{d}{dt}\ln{\left(\varepsilon_{||}\mu_{\bot}\right)} \,,
$$
\begin{equation} \label{AP33}
\Pi^{1(BA)}_{1t}= \Pi^{2(B)}_{2t} = -
\frac{1}{2}\frac{d}{dt}\ln{\left(\varepsilon_{\bot}\mu_{||}\right)}
\,, \quad \Pi^{3(A)}_{3t}= \Pi^{3(B)}_{3t} = -
\frac{1}{2}\frac{d}{dt}\ln{\left(\varepsilon_{\bot}\mu_{\bot}\right)}
\,,
\end{equation}
where the quantities $\varepsilon_{||}$, $\varepsilon_{\bot}$,
$\mu_{||}$ and $\mu_{\bot}$ are defined in (\ref{Bepsilon}) and
(\ref{Bmu}) with the susceptibility tensor components from
(\ref{R1})-(\ref{R5}). Thus, the force $F^{i (\alpha)}$ consists
of derivatives of $a(t)$ and $c(t)$ up to the third order.


\begin{acknowledgments}
The authors are grateful to Prof. W. Zimdahl for the fruitful
discussion. This work was supported by the Deutsche
Forschungsgemeinschaft through the project No. 436RUS113/487/0-5, and partially
by the Russian Foundation for Basic Research through the grant No 08-02-00325-a.
\end{acknowledgments}


\end{document}